\documentclass[12pt]{article}

\usepackage{epsfig}

\usepackage{amssymb}
\usepackage{amsfonts}

\usepackage{color}
\def\be{\begin{equation}}
\def\ee{\end{equation}}
\def\ba{\begin{array}{c}}
\def\ea{\end{array}}

\newcommand{\bea}{\begin{eqnarray}}
\newcommand{\eea}{\end{eqnarray}}

\newcommand{\kt}{\rangle}

\newtheorem{thm}{Theorem}
\newtheorem{cor}[thm]{Corollary}
\newtheorem{lemma}[thm]{Lemma}

\begin{document}

\begin{center}

{\Large \bf

Interference of non-Hermiticity with Hermiticity
at exceptional points

}

\vspace{0.8cm}

  {\bf Miloslav Znojil}$^{1,2}$

\end{center}

%
%
%\vspace{1mm} Nuclear Physics Institute of the CAS, Hlavn\'{\i} 130,
%250 68 \v{R}e\v{z}, Czech Republic
%
%

\vspace{10mm}

 $^{1}$ {Department of Physics, Faculty of
Science, University of Hradec Kr\'{a}lov\'{e}, Rokitansk\'{e}ho 62,
50003 Hradec Kr\'{a}lov\'{e},
 Czech Republic}

 $^{2}$ {The Czech Academy of Sciences,
 Nuclear Physics Institute,
 Hlavn\'{\i} 130,
250 68 \v{R}e\v{z}, Czech Republic, {e-mail: znojil@ujf.cas.cz}}

 %\footnote{{e-mail: znojil@ujf.cas.cz}}

%\newpage

\vspace{10mm}

\section*{Abstract}

A ``solvable'', i.e., partially non-numerically
tractable
family of quantum $N$ by $N$ matrix
Hamiltonians mixing Hermiticity with
non-Hermiticity is constructed and studied
at any even $N=2J$.
These operators are shown to support the existence
of the Kato's exceptional-point (EP) degeneracies.
A dedicated perturbation theory
is recalled to show that
under a generic perturbation $\lambda\,V$
the ``unfolding'' removal of the
(in general, up to $N-$fold)
EP degeneracy
proceeds via a complex energy spectrum.

\newpage

\section{Introduction}

In the conventional quantum mechanics of textbooks \cite{Messiah}
the unitarity of evolution is interpreted as
a consequence of the self-adjointness
of the Hamiltonian
in Schr\"{o}dinger picture \cite{Stone}.
A
``quasi-Hermitian operator'' \cite{Dieudonne}
modification of this interpretation
of Schr\"{o}dinger picture
has been offered, in 1992,
by Scholtz et al \cite{Geyer}. These
authors characterized the conventional
mathematical interpretation
of the unitarity of the evolution as
``somewhat restrictive''.
Assuming that
in a suitable preselected ``mathematical'' Hilbert space ${\cal H}_{math}$
a bound-state Hamiltonian $H$
appears
non-Hermitian, it has been argued that
the system in question can still acquire
``the normal quantum mechanical interpretation'' in another,
``physical'' Hilbert space ${\cal H}_{phys}$
endowed with an amended, Hamiltonian-Hermitizing inner-product
metric $\Theta =\Theta(H) \neq I$.

The idea
of the Hermitization of
the operators of observables
via an
{\it ad hoc\,} amendment
of the inner product
in ${\cal H}_{phys}$
is currently widely used by physicists
(see, e.g., the recent reviews \cite{Carl,ali,Carlbook}).
At the same time,
it deeper analyses by mathematicians
(see, e.g.,  \cite{Trefethen,book,Viola})
reveal that the formalism
(to be called here quasi-Hermitian quantum mechanics, QHQM)
still requires a more rigorous critical re-evaluation
of the conditions and boundaries of its applicability.

The doubts and critical voices
-- which in fact motivated also our present study --
appeared also among physicists.
In 2007, in particular,
Hugh Jones
studied the problem of scattering \cite{Joneskonf,Jones} and
he noticed
that in certain quantum systems characterized
by the coexistence,
mixture
and interaction
between a non-Hermitian and a Hermitian
Hamiltonian
the QHQM theory
of scattering
may really become conceptually inconsistent.
Incidentally, what followed in the research was ``the
weakening of emphasis on the explicit
constructions of metrics'' \cite{MZbook}.
The crisis emerged but ``it did not hit the models using
effective Hamiltonians'' for which
``their phenomenological interpretation
does not require any construction of the metric''
\cite{MZbook}.

A mathematical consistency of the QHQM theory
of unitary systems can still
be achieved
after its more consequent formulation
\cite{Geyer,ali,book}.
The price to pay is that
the
choice (or rather the construction) of the
internally consistent pairs of a Hamiltonian $H$
(which is admitted non-Hermitian in ${\cal H}_{math}$)
and of the {\it ad hoc\,} Hermitizing
inner-product metric $\Theta(H)$
(which must be Hermitian in ${\cal H}_{math}$)
is technically
difficult, especially, as emphasized by Jones \cite{Jones},
in the scattering dynamical regime.
In the literature an escape out of the dilemma
has been found in an {\it ad hoc\,} restriction
of the admissible non-Hermitian Hamiltonians,
say, to their bounded-operator subclass \cite{Geyer},
or to their strictly confining bound-states-generating forms
\cite{BJones}.
Along these lines it has been found that
the
technical aspects of the analysis of the non-Hermitian
models appear perceivably less complicated
in the
narrower unitary
bound-state dynamical context
than in the full-fledged quantum theory
including scattering
(see, nevertheless, a few fully consistent
QHQM models of scattering in \cite{scatt,discrete}).
Alternatively, attention has been turned to
the
study of the open, non-unitary quantum systems
with resonances \cite{Nimrod}.

In both the unitary and non-unitary contexts,
several
serendipitious phenomenological benefits
have been found during the
discussions of the role of the
so called Kato's exceptional points (EP, \cite{Kato}).
In this respect, for the time being,
let us only remind the readers that even after
the restriction of attention to the Hermitizable
Hamiltonians with the
discrete and real spectra, it has been found that
the QHQM simulation of the dynamics
using the nontrivial metrics $\Theta$
becomes much more flexible
than its traditional Hermitian forms in which
the metric remains fixed and trivial, $\Theta=I$,

The existence of the new model-building freedom
using the non-Hermiticity
opens new horizons not only
via the inventions of the pragmatically oriented
phenomenological models (say, of
quantum phase transitions \cite{denis,
denisb,denisc,denisd,denise,denisf,denisg,denish,PRSA})
but also via some more formal mathematical toy-model
constructions
(cf., e.g.,  \cite{Uwe,Uweb,catast}).
In both the open and closed quantum systems
one even encounters and identifies
various genuine quantum analogues of
what is called, in non-quantum physics,
evolution bifurcations or, in the
Thom-inspired terminology,
``catastrophes'' \cite{Zeeman,Zeemanb}.
In all of these contexts the key role is played, indeed,
by the above-mentioned concept of the
Kato's exceptional points.
In what follows, this subject is
further to be developed.

For methodical purposes, the form of our
argumentation and
toy models
will be chosen as simple as possible.
In fact,
our study of
interaction between the
comparatively well separated
Hermitian and non-Hermitian components of the Hamiltonian
will be essentially simplified
by not involving the systems with scattering, and
by the
technically motivated
restriction of our analysis to certain
EP-related phenomena supported by the
$N$ by $N$
matrix forms of $H=H^{(N)}$ with even $N=2J$.
One of the key merits of our results
and conclusions will be that they
will be mostly algebraic and non-numerical
even when
the matrix size $N$ itself will be admitted
arbitrarily large.

\section{Non-Hermitian and Hermitian operators
in interaction}

The
key appeal of the use of the quantum
non-Hermitian Hamiltonians
is that they admit access to the
Kato's exceptional-point (EP, \cite{Kato})
dynamical singularities.
In particular, in
some special unitary QHQM systems living in the vicinity
of these singularities
the non-Hermiticity might help to
mimic, e.g., the quantum phase transitions \cite{denis,PRSA}.
Still, on both the mathematical
and phenomenological sides the subject is full of open questions.
In this framework we intend to address,
first of all, those aspects of the
generic Hermitian -- non-Hermitian interference
theory which are related to the Jones'
ansatz
 \be
 H = H_1+H_2+H_{coupling}\,,\ \ \ \
 H_1=H_1^\dagger\,,\ \ \ \
 H_2 \neq H_2^\dagger\,.
 \label{quando}
 \ee
We will develop a class of models elucidating some of
the general phenomenological
as well as mathematical
consequences of this ansatz.
In addition to the mathematical orientation of our present study
we will also stress the existence of a parallel
motivation of our project of the study of
the interactions (\ref{quando}) and of the
EPs in physics.

\subsection{Motivation: the access to EPs in quantum physics}

For any parameter-dependent family of
non-Hermitian quantum
Hamiltonians $H=H(\lambda)$ in (\ref{quando})
one can expect
the existence of a ``critical value'' of the
parameter at which the spectrum ceases to be real
\cite{BJones}. In the language of mathematics such a
``critical value'' of the coupling
is to be identified with the Kato's EP parameter $\lambda^{(EP)}$.
Recently, the concept
acquired an immediate experimental meaning
in several phenomenological applications ranging from the
relativistic quantum mechanics \cite{alliKG,jaKG} and
quantum cosmology \cite{aliWDW,WDW}
up to the efficient toy-model
simulations of the various forms of quantum
phase transitions \cite{Denis,Denisb,Denisc}.

Many years ago, the use of the EPs already caused
a change of the paradigm
in the mathematical foundations of the
perturbation theory of linear operators \cite{Kato,Simon}.
Still, until recently,
the concept did not seem to have found an immediate
applications in
the conventional
textbook quantum physics \cite{Messiah}.
Indeed, in the context of the theoretical
quantum physics of unitary systems
the typical EP singularities
were complex so that their
experimental visibility
remained, necessarily, indirect.

In his old but still fully authoritative monograph \cite{Kato},
Kato emphasized that for a given
self-adjoint Hamiltonian
$\mathfrak{h}(\lambda)=\mathfrak{h}^\dagger(\lambda)$
which is analytic in the parameter,
our knowledge of the (complex) quantities $\lambda^{(EP)}$
opens the way towards a rigorous
determination of the radius of convergence $R$
of the most common Rayleigh-Schr\"{o}dinger perturbation series for
the bound state energies $E_n(\lambda)$.
This leads to a paradox that
in one of the most prominent
examples of the applicability of the
Rayleigh-Schr\"{o}dinger perturbation series
to the quartic anharmonic quantum oscillator
$\mathfrak{h}^{(AHO)}(\lambda)=p^2+\lambda\,q^4$
the authors of paper \cite{BenderWu}
(who managed to localize
practically all of the not too large
values of $\lambda^{(EP/AHO)}_{(m,n)}$
(numbered by the two non-negative integers) numerically)
came to the disappointing conclusion that $R^{(AHO)}=0$.

In the language of experimental physics
an analogous discouraging conclusion is that
every EP value $\lambda^{(EP)}$ of any variable parameter
entering any operator representing an observable quantity must
necessarily remain
unphysical because
${\rm Im\,}\lambda^{(EP)} \neq 0$.
Thus, in experiments, even the EP singularity
(with a sufficiently small size of
its imaginary part ${\rm Im\,}\lambda^{(EP)}$)
only manifest their presence via the well known
phenomenon of the so called avoided level crossing.

During the recent developments
of the study of EPs, fortunately, several
``dark clouds'' (i.e., e.g., the above-mentioned and
rather unpleasant rigorous proof of the
manifest divergence of the AHO series)
appeared to have also a silver linen
(in the context of perturbation theory, for example,
several alternative,
sophisticated but unexpectedly efficient
resummation methods have been discovered
as a consequence \cite{Simon}).
The subsequent
turn of attention to the analytic continuation methods
resulted, finally, in the fairly unexpected Bender's and Boettcher's
discovery \cite{BB}
that in some quantum models
characterized by an analytically continued and
manifestly non-Hermitian Hamiltonian
$H(\lambda) \neq H^\dagger(\lambda)$
the bound state energies $E_n(\lambda)$
(as well as at least some of the related
EP values $\lambda^{(EP)}_{(m,n)}$ of the parameter)
still
might
keep their traditional experimental bound-state meaning while being
real, discrete and bounded from below.
On these grounds,
Bender with Boettcher \cite{BB} conjectured that there might exist
certain manifestly non-Hermitian
realizations of quantum systems
which could still admit
the conventional probabilistic interpretation and properties.
Their claim has been illustrated
by the manifestly non-Hermitian ordinary differential Hamiltonians
 \be
 H^{(BB)}(\lambda)=-\frac{d^2}{dx^2} + g(\lambda,x)\,x^2\,,
 \ \ \ \
 g(\lambda,x)= ({\rm i}x)^\lambda\,,
 \ \ \ \ x \in (-\infty,\infty)\,.
 \label{haber}
 \ee
which
``may be viewed as analytic continuations of
conventional theories'' \cite{BB}.
In the limit $\lambda\to 0$, indeed, their toy
model $H^{(BB)}(\lambda)$ coincides with
the standard self-adjoint harmonic oscillator with
equidistant spectrum  $E^{(BB)}_n(0)=2n+1$.

%\section{The concept of hiddenly Hermitian observables}

Equation (\ref{haber})
represents one of the most popular examples
of an operator
which is manifestly non-Hermitian (in $L^2(\mathbb{R})$) but still
having spectrum which is real, discrete and bounded from below \cite{BB}.
Thus, it has long been believed that
such an operator
might represent a more or less standard quantum
bound-state Hamiltonian
``which allows for the normal
quantum-mechanical interpretation'' \cite{Geyer}.
Unfortunately, a deeper analysis of the operator
revealed that these expectations cannot be fulfilled \cite{Siegl}.
In a partial analogy with the
above-mentioned $R^{(AHO)}=0$ paradox,
the limiting transition $H^{(BB)}(\lambda) \to H^{(BB)}(0) $
has to be interpreted as discontinuous. Indeed,
even the smallest decrease of the exponent $\lambda$
below zero leads to an abrupt loss of the reality
of the whole high-energy part of the spectrum.
Only
a few low-lying levels
$E^{(BB)}_n(\lambda)$ with $n=0,1,\ldots,n_{\max}(\lambda)$
remain real, with the upper bound integer
$n_{\max}(\lambda)$ growing,
as a function of $\lambda \in (-2,2)$,
between its minimum $n_{\max}(\lambda_{\min})=0$ at $\lambda_{\min} \in (-2,-0.57793)$
and its
maximum $n_{\max}(\lambda_{\max})=\infty$, say, in the
interval of $\lambda_{\max} \in [0,2)$.

With $\lambda_{\max} \in [0,2)$ people still felt inclined to treat
the evolution generated by
Hamiltonian (\ref{haber})
as one of the most impressive (i.e., local-interaction-controlled)
benchmark examples of the quantum evolution
which is ``hiddenly'' unitary. In brief
the widely accepted belief was that
the hidden unitarity {\it alias\,} quasi-unitarity
of the model
is
equivalent to the conventional
unitarity \cite{Geyer,ali}).
In connection with the
local-interaction-controlled family of models (\ref{haber}),
unfortunately,
the latter optimistic expectations remained unfulfilled:
In 2012, Siegl with Krej\v{c}i\v{r}\'{\i}k
considered their special case
with $\lambda=1$ \cite{Siegl}, and they proved, using the rigorous
methods of functional analysis, that
``there is no quantum-mechanical Hamiltonian
associated with it''
(i.e., with the cubic-interaction
BB operator $H^{(BB)}(\lambda)$ at $\lambda=1$) \cite{Siegl}.
Moreover, these authors also added that ``the method \ldots of the
disproval \ldots
does not restrict to the particular Hamiltonian (\ref{haber}).
It also applies \ldots to many others'' \cite{Siegl}.

In such a situation, obviously, it was necessary to turn attention
to the other sufficiently elementary toy models.

\subsection{Paradox of non-locality of complex delta-function interactions}

In our present paper the latter words of warning
will be interpreted in three ways.
Firstly, we will accept the fact that the Siegl's and Krej\v{c}i\v{r}\'{\i}k's
disproof of the quasi-Hermiticity of the Bender's and Boettcher's
(or, more precisely, of the
Caliceti's \cite{Caliceti}
and Bessis' \cite{DB}
or Alvarez's \cite{BenderWub})
imaginary cubic benchmark example
is correct. After all, their observations appeared
reconfirmed
in
a few other studies \cite{Viola,Uwe,Uweb}.
Secondly, we are persuaded that the criticism of the specific model(s)
does
not imply a disproval of the applicability and appeal
of the abstract quantum mechanics
in its innovative quasi-Hermitian formulation as provided, say,
by reviews \cite{Geyer,ali,SIGMA}.
Hence, we believe that the alternative, still sufficiently transparent
benchmark models have to be sought.
Last but not least, thirdly,
we are going to oppose the methodical scepticism
connected with the formalism and
formulated, recently, in several independent studies \cite{ozky}.
Thus, in our present paper we
will express
and constructively support
our belief that
the quasi-Hermitian
quantum theories, ``originally proposed with the aim of extending
standard quantum theory by relaxing the Hermiticity
constraint on Hamiltonians'' \cite{ozky},
really {\em did\,} reach the goal.

In the dynamical scenarios characterized by the
interference between
Hermitian and non-Hermitian
components in the Hamiltonian (\ref{quando})
the theoretician may encounter ``a quandary'' \cite{Joneskonf}.
The picture of physics may easily become deeply unsatisfactory
in {\em both\,} of the alternative representation Hilbert spaces
${\cal H}_{math}$ and ${\cal H}_{phys}$.
A more detailed description of the dilemma has been
offered by Jones \cite{Jonesb}. In his paper
the author studied a hypothetical scattering controlled
by the delta-function interactions in Hamiltonian
 $
 H=H_1(\alpha)+H_2(\beta,L)
 $
where
 \be
 H_1(\alpha)= -\frac{\hbar^2}{2m}\,\frac{d^2}{dx^2}-\alpha\,\delta(x)
 =H_1^\dagger(\alpha)\,
 \ee
and
 \be
 H_2(\beta,L)=-\frac{\hbar^2}{2m}\,\frac{d^2}{dx^2}+ {\rm i}\beta\,
 \left [\delta(x-L)\-\delta(x+L)
 \right ]\neq H_2^\dagger(\beta,L)\,.
 \label{dellt}
 \ee
In the conventional Hilbert space ${\cal H}_{math}$
the first component
of the Hamiltonian was Hermitian
but its second part has been chosen non-Hermitian, i.e.
all of the three independent
parameters $\alpha$, $\beta$ and $L$ were chosen real.
After a detailed analysis
based on perturbation theory
the Jones' conclusion was
that ``the conceptual issues arise'' \cite{Jones,Jonesb}.

In  \cite{Jonesb}, in particular, we read that
for the scattering states  ``one can \ldots no longer talk in terms
of reflection and transmission coefficients'', and that ``the only
satisfactory resolution of this dilemma
is to treat the non-Hermitian scattering potential as an effective one,
\ldots accepting that this
\ldots may well involve the loss of unitarity'' \cite{Jonesb}.

The puzzle found its resolution in
\cite{scatt,smeared}.
It has been revealed there that in the
slightly counterintuitive QHQM framework
a very careful experimental as well as
theoretical definition of
the scattering setup is obligatory.
More specifically,
the source of the apparent inconsistencies has been found to lie
in an inappropriate nature of the
non-Hermitian scatterer
$H_2$ of Eq.~(\ref{dellt}) because
its spatial range appeared infinite
in the physical Hilbert space ${\cal H}_{phys}$.
Thus, paradoxically but understandably,
the phenomenologically natural physical
finite-range requirement
excludes the delta-function model of Eq.~(\ref{dellt})
as unacceptable.

\section{Finite-dimensional toy-model Hamiltonians}

%\subsection{Exceptional points}block-tridiagonal

%Toy model (\ref{mohs}) with $N=8$

The ill-conditioned nature of the numerical
search for the EPs in general cases
\cite{sqwEP}
was one of the reasons why
we choose,
in our present toy model Hamiltonians (\ref{mohs}),
the equidistant main diagonal.
Indeed, in the light of our older paper \cite{maximal},
such a choice can be perceived as technically important,
leading to an enormous simplification
of the process of the localization of its
EP-related extreme-physics limits.
Needless to add that
by the needs of phenomenology
such a localization is very well motivated
because in the models of some
fairly realistic quantum systems,
the values of the EPs become related to the
instants of a genuine quantum phase
transition \cite{PRSA}.

%Block-tridiagonality requirement

\subsection{Partitioned structure of the relevant Hamiltonians}

In the realistic models
connecting a Hermitian and non-Hermitian
components in an operator of an observable
quantity
the study
of the mutual %$H_1 \leftrightarrow H_2$
interference of the components
is difficult.
The mathematical methods of the
corresponding physics-oriented
analysis of the
system's properties are usually  approximate.
For illustration we may recall the
Bender's and Jones'
paper \cite{BJones} in which they considered
several ``realistic'', differential-operator models
which really appeared tractable
by the conventional perturbation theory
in its leading-order form.

In {\it loc. cit.} the authors also admitted that
the insight
in the features of the scattering models is
less easy. At the same time, they reconfirmed
the
conceptual consistency of the QHQM approach
to the
class of the mixed, ``composite'', three-component bound-state
Hamiltonians (\ref{quando})
``by means of a number of soluble models''
\cite{BJones}.
These authors have shown that in all of their models
``the energy remains real
for small values of the coupling constant''
in $H_{coupling}$ \cite{BJones}.
On the methodical level these
conclusions
were further supported by
the four-by-four
complex-matrix
sample of superposition~(\ref{quando}).

The latter argument did not sound too persuasive.
Its authors
treated even their elementary small-matrix model
by the mere brute-force numerical
diagonalization.
Indirectly, the approximate nature of
the result
contributed to the motivation of
our present paper. We intend to complement
the analysis
of the Hermiticity -- non-Hermiticity interference
and interfaces by its extension to
a family of the
bound-state models which would be solvable exactly.
We will restrict our attention to the specific
real-matrix Hamiltonians characterized by
an arbitrary finite (and, for the sake of definiteness,
even) matrix dimension $N=2M+2K$.
In this manner we will complement the
above-mentioned studies based on
the mere approximate estimates.

More specifically,
our present toy model Hamiltonians will all have the
block-tridiagonal partitioned-matrix structure
 \be
 H=H^{(N)}=
 \left[ \begin {array}{c|c|c}
 A&U_+&0\\
 \hline
 U_-&C&V_+
 \\
 \hline
 0&V_-&B
 \end {array} \right]\,.
 \label{speci}
 \ee
The two $M$ by $M$ submatrices
$A=A^\dagger$ and $B=B^\dagger$ will define the
block-diagonal
Hermitian sub-Hamiltonian component $$H_1=
 \left[ \begin {array}{c|c|c}
 A&0&0\\
 \hline
 0&0&0
 \\
 \hline
 0&0&B
 \end {array} \right]
=H_1^\dagger.$$
The complementary non-Hermitian
component $H_2\neq H_2^\dagger$ will only
contain the single
non-vanishing $2K$ by $2K$
submatrix $C\neq C^\dagger$. The remaining
four non-vanishing submatrices $U_\pm$ and $V_\pm$
will finally form $H_{coupling}$
controlling, in (\ref{quando}), the interaction
between the latter two sub-Hamiltonians.

\subsection{${\cal PT}-$symmetry requirement and reparametrization}

The main goal of our construction of the benchmark models (\ref{speci})
will be twofold. Firstly, we will search for the models in which
the integers $M$ and $K$ (specifying the submatrix dimensions)
can be arbitrary.
A wealth of the
phenomenologically interesting
spectral structures may be then expected to occur.
Secondly, we will
succeed in keeping our study non-numerical. We will
obtain and describe a family of benchmark Hamiltonians
for which the demonstration of the reality of the spectrum and/or
of its loss will be algebraic and rigorous.

We will see that
the unitary, closed-system evolution regime
characterized by the reality of spectrum
need not
necessarily require the smallness of the couplings.
In the
mathematically not less interesting
complex-spectrum scenarios
the evolution proceeds in the open-system regime.
One  changes the philosophy and,
in the words of Ref.~\cite{Jonesb},
``accepts that unitarity is not conserved, essentially because
we are dealing with
a subsystem of a larger system whose physics
has not been taken fully into
account''.

In the methodical framework,
our present choice of a specific
realization of the family of Hamiltonians (\ref{speci})
will be restricted by the following three requirements.

\begin{itemize}

\item {[A]}
the elements of $H$ lying on the main diagonal will
form a real and equidistant sequence simulating the
spectrum of the most common
harmonic oscillator;

\item {[B]}
as long as
a broad class of general matrices can be routinely
tridiagonalized,
we will assume that all of our $H$s are tridiagonal.
Moreover, for the methodical reasons formulated in \cite{tridiagonal},
their off-diagonal part will be chosen antisymmetric;

\item {[C]}
after a finite-dimensional truncation,
our benchmark $N$ by $N$ matrices $H=H^{(N)}$
will be required ${\cal PT}-$symmetric,
$H{\cal PT} = {\cal PT}H$ \cite{BG,AKbook}.
Here,
${\cal P}$
is defined as
the antidiagonal unit matrix (``parity'' \cite{tridiagonal}) while
the symbol ${\cal T}$
represents the antilinear Hermitian
conjugation mimicking the time
reversal \cite{BB}.

\end{itemize}

 \noindent
In the light of the first two requirements [A] and [B] we may
start our considerations from the anharmonic-oscillator-like
infinite-dimensional
tridiagonal matrix
 \be
 H^{(\infty)}=
 \left[ \begin {array}{ccccc}
 1&{ a_1}&0&0&\ldots\\
 \noalign{\medskip}-{ a_1}&3&{ a_2}&0&\ldots
 \\
 \noalign{\medskip}0&-{ a_2}&5&{ a_3}&\ddots
 \\
 \noalign{\medskip}0&0&-{ a_3}&7&\ddots
 \\
 \noalign{\medskip}\vdots
 &\vdots&\ddots&\ddots&\ddots
 \end {array} \right]\,.
 \label{concon}
 \ee
Such an
an infinite-dimensional matrix Hamiltonian is composed of
the main diagonal (which represents the equidistant energy spectrum of
an unperturbed harmonic oscillator) and of the off-diagonal
``perturbation'' having the maximally elementary
form of a real and antisymmetric two-diagonal
multi-parametric anharmonic interaction,
$H^{(\infty)}=H^{(HO)}+ V(a_1,a_2,\ldots)$.

In
a way well motivated, say, in the context of the numerical and/or
variational calculations
one usually truncates
the infinite-dimensional
matrix (\ref{concon}) to its finite-dimensional $N$ by $N$ matrix
alternative. Then,
we may trivially replace
$H_{nn} \to H_{nn} - constant$
(i.e.,  shift the
conventional origin of the energy scale).
We may also impose the
${\cal PT}-$symmetry requirement [C].
All this will lead to our ultimate replacement of Eq.~(\ref{concon})
by the matrix
 \be
 H^{(2J)}=%H^{(HO)}+ V(a_1,\ldots,a_{J})=
 \left[ \begin {array}{cccccccc}
 1-2J&{ a_1}&0&0&\ldots&\ldots&0&0\\
 \noalign{\medskip}-{ a_1}&3-2J&{ a_2}&0&0&\ldots&\ldots&0
 \\
 \noalign{\medskip}0 &-{ a_2}&\ddots&\ddots&\ddots&\ddots&&\vdots
 \\
 \noalign{\medskip}0&\ddots&\ddots&-1&{ a_J}&0&\ddots&\vdots
 \\
 \noalign{\medskip}\vdots&\ddots&0&-{ a_J}&1&\ddots&\ddots&0
 \\
 \noalign{\medskip}\vdots&&\ddots&\ddots&\ddots&\ddots&{ a_2}&0
 \\
 \noalign{\medskip}0&\ldots&\ldots&0&0&-{ a_2}&2J-3&{ a_1}
 \\
 \noalign{\medskip}0&0&\ldots&\ldots&0&0&-{ a_1}&2J-1
 \end {array} \right]\,.
 \label{umohs}
 \ee
We have to distinguish between the
parameters $a_j$ which are
real
(for them, the measure of non-Hermiticity  $(-a_j)-(a_j^\star) $ is maximal)
and which are purely imaginary
(for them, the Hermiticity is guaranteed since  $(-a_j)-(a_j^\star) =0$).
For this reason it makes sense to reparametrize
  \be
  a_j=\sqrt{{z}_j}\,,\ \ \ \ j = 1,2,\ldots,J\,.
  \label{repa}
  \ee
In what follows
we will
simplify the situation
by an additional ordering assumption
 \be
 {z}_1 < {z}_2 < \ldots < {z}_J \,.
 \label{ordered}
 \ee
This will reduce the menu of the
eligible partitionings
of
our Hamiltonians $H^{(N)}$
to the mere triply partitioned
pattern of Eq.~(\ref{speci}).
In this equation, due to the assumption of the ${\cal PT}-$symmetry of $H^{(N)}$,
one merely has to consider the
interaction between the
two Hermitian $M$ by $M$ submatrices $A$ and $B$
and the single non-Hermitian $2K$ by $2K$ submatrix $C$.
At a fixed $N=2J=2M+2K$, naturally, we only have a single
variable dimension (i.e., $M$ or $K$) at our disposal,
the value of which depends on the parameters $z_j$
in a way classified in Table~\ref{Pwe1}.

\begin{table}[h]
\caption{Dimensions in partitioning
(\ref{speci})
under assumption (\ref{ordered}).}
 \label{Pwe1}
\centering
\begin{tabular}{||c||c|c|c|c|c|c||}
\hline \hline
 &${z}_1>0$
 &${z}_1\leq 0$
 &${z}_2\leq 0$
 &\ldots
 &${z}_{J-1}\leq 0$
 &$-$
 %&${z}_J<0$
 \\
 &$-$
 &${z}_2> 0$
 &${z}_3> 0$
 &\ldots
 &${z}_{J}> 0$
 &${z}_J= 0$
 %&${z}_J<0$
 \\
 \hline \hline
 $M$
 &$-$
 &$1$
 &$2$
 &\ldots
 &$J-1$
 &$J$
 %&-
 \\
 \hline \hline
 $K$
 &$J$
 &$J-1$
 &$J-2$
 &\ldots
 &$1$
 &$-$
 %&-
 \\
 \hline \hline
\end{tabular}
\end{table}

\section{Exceptional points}

In the mathematically oriented context of our present paper
let us point out
that given a
single-parameter-dependent tridiagonal matrix $H=H(\lambda)$,
the search for its EP singularities $\lambda^{(EP)}$
is difficult even if we
keep the main diagonal parameter-independent.

For an illustration of the specific difficulties emerging
in the non-equidistant diagonal-element cases
we recommend
the details of the EP construction as described in \cite{sqwEP}.
It has been shown there that the straightforward,
brute-force localization of the values of $\lambda =\lambda^{(EP)}$
is truly an extremely ill-conditioned numerical task in general.
In parallel,
from time to time,
we will also
simplify our argumentation
by a return
from the fully general dimension $N=2J$
to its various comprehensive
exemplifications.

\subsection{Illustrative $N=8$ example}

%\subsection{$N=8$ illustration}

Without the
assumption of the ${\cal PT}-$symmetry
of $H(\lambda)$,
even the equidistance of the elements on the main diagonal
wouldn't be enough, leaving still
the
printed version of the secular polynomial
 $P(E)=\det (H-{E}\,I)$
(evaluated via the
computer-assisted symbolic manipulations in MAPLE \cite{Maple})
prohibitively long.
Fortunately, after the change of variables (\ref{repa})
we managed to
shorten the printout of $P(E)$ to an acceptable length.

The mere four rows were needed at
the not too small $N=8$.
For this reason, such a choice is also suitable
for our present methodical purposes since
the structure of our Hamiltonians (\ref{umohs})
is
well sampled by the model with $J=4$,
 \be
 H^{(8)}=H^{(8)}(a_1,\ldots,a_4)=
\left[ \begin {array}{cccccccc} -7&{ a_1}&0&0&0&0&0&0\\
\noalign{\medskip}-{ a_1}&-5&{ a_2}&0&0&0&0&0\\\noalign{\medskip}0
&-{ a_2}&-3&{ a_3}&0&0&0&0\\\noalign{\medskip}0&0&-{ a_3}&-1&{
a_4}&0&0&0\\\noalign{\medskip}0&0&0&-{ a_4}&1&{ a_3}&0&0
\\\noalign{\medskip}0&0&0&0&-{ a_3}&3&{ a_2}&0
\\\noalign{\medskip}0&0&0&0&0&-{ a_2}&5&{ a_1}
\\\noalign{\medskip}0&0&0&0&0&0&-{ a_1}&7\end {array} \right]\,.
 \label{mohs}
 \ee
As long as every submatrix
of $H_{coupling}$
(i.e., $U_\pm$ and $V_\pm$ in (\ref{speci}))
will have, by construction,  a single non-vanishing matrix element,
the above-introduced change of variables $a_j \to {z}_j$
enables us to switch, easily, between the
Hermitian and non-Hermitian
form of the coupling matrix
itself. At a fixed
$N=8$ it makes sense to
avoid the
need of subscripts and write
 \be
 a_1:=\sqrt{D};
 \ \ \ \
 a_2:=\sqrt{C};
 \ \ \ \
 a_3:=\sqrt{B};
 \ \ \ \
 a_4:=\sqrt{A}\,.
 \label{ctyri}
  \ee
The use of a fixed $N$ will
shorten the necessary
compact review of some older relevant results in the literature.
Among the most relevant ones let us mention the descriptions
of the less standard QHQM formalism.
Although its short summary may be found in Appendix A below,
it is necessary to remind the readers that
one of the key {\em practical\,} benefits
offered by such a formalism
lies in the possibility of the description
of the various (i.e., both closed and open) quantum systems
in an arbitrarily small vicinity of
their exceptional-point (EP, \cite{Kato}) degenerate extremes.

This has both its mathematical and physical aspects,
both of which may be well illustrated by the specific choice of $N=8$.
Using reparametrization (\ref{ctyri})
we obtain the reasonably printable secular polynomial
 \be
 \det (H-{E}\,I)=
{{ {E}}}^{8}+ f_6\, {{ {E}}}^{6}+
 f_4\,{{ {E}}}^{4}
 + f_2\,{{ {E}}}^{2}+f_0
 \ee
where $f_6={ -84+2\,D+2\,C+2\,B+A }$ while
  $$
  f_4=
  2\,C { D } +50\,D+4\,B { D } +2\,A
 { D } +{D}^{2}-70\,C+1974-142\,B-83\,A+2\,BC+2\,AC+{C}^{2}
+{B}^{2}
$$
and
 $$
 f_2=
 -12916-682\,D-74\,{B}^{2}+2\,{
B}^{2} { D } +2006\,B-1402\,C-50\,{C}^{2}-44\,C { D
 } +2\,CB { D } +
 $$
 $$
 +2\,CA { D } +A{C}^{2}-68
\,AC+52\,A { D } +1891\,A+152\,B { D } -108\,BC+
2\,B{D}^{2}+A{D}^{2}-10\,{D}^{2}
 $$
and
 $$
 f_0=
 11025+630\,D+
1225\,{B}^{2}+70\,{B}^{2} { D } +7350\,B+1470\,C+49\,{C}^{2
}+42\,C { D } +
$$
$$
+
14\,CB { D } -42\,CA { D
 } -49\,A{C}^{2}-1470\,AC-630\,A { D } -11025\,A+420
\,B { D } +490\,BC+
$$
$$
+
6\,B{D}^{2}-9\,A{D}^{2}+9\,{D}^{2}+{B}^{
2}{D}^{2}\,.
 $$
The search for the eightfold EP degeneracy can be found performed
in \cite{maximal}.
It yielded
the following
result, with the proof described
in \cite{maximal}.

\begin{lemma} \cite{maximal}
\label{lemmamax}
For our toy model (\ref{mohs}) + (\ref{ctyri}), four equations
 \be
 f_0(A,B,C,D)=0\,,\ \ \
 f_2(A,B,C,D)=0\,,\ \ \
 f_4(A,B,C,D)=0\,,\ \ \
 f_6(A,B,C,D)=0\,
 \ee
representing
the necessary condition of the existence of
the EP singularity of order eight (=EP(8))
specify a
unique and exact quadruplet of the positive
values of the parameters
 \be
 A=A_0=16,\ \ \ \
 B=B_0=15,\ \ \ \
 C=C_0=12,\ \ \ \
 D=D_0=7\,.
 \label{onyctyri}
 \ee
\end{lemma}
In {\it loc. cit.\,} the latter
result was characterized as,
from two separate points of view, surprising.

\subsection{Hamiltonians in the EP limit}

The first surprise was that for
Hamiltonian (\ref{mohs}) the calculation led,
in the
EP limit, to a
maximal EP(8) degeneracy,
i.e., to the maximal dimension $N=2K$ and minimal $M=0$ in the
partitioned form~(\ref{speci}) of the Hamiltonian.
The second surprise was, perhaps, even more impressive
because after an extension of the analysis to the general,
$N$ by $N$ matrix
analogue of matrix (\ref{mohs})
(with the odd$-N$ results added in \cite{tridiagonal}),
all of the off-diagonal matrix elements supporting
the analogous EP(N) degeneracy
appeared to have an analogous, strictly non-numerical form.

\begin{lemma} \cite{maximal,tridiagonal}
\label{lemmaNmax}
For the general $N$ by $N$
matrices (\ref{umohs}),
the necessary condition of the existence of
the EP singularity of order $N$ (=EP(N))
appeared to have the
unique and exact solution
 \be
 z_k^{(EP(N))}={k(N-k)}\,,\ \ \ \ k = 1,2,\ldots,[N/2]\,.
 \label{papa}
 \ee
\end{lemma}
Thus, in particular, at any even $N=2J$ the
values of all of the
off-diagonal matrix elements in the
EP(N) limit of the Hamiltonian are known exactly.

For the sake of brevity we do not intend to
consider any
models possessing
the infinite dimension $N=\infty$
because multiple technical subtleties of the
formalism of functional analysis would have
to be necessarily added.
At any finite $N<\infty$, in contrast, it is
very easily seen that
the standard physical probabilistic interpretation of
the system is lost
in the EP limit  $H \to H^{(EP)}$.
From the point of view of mathematics, nevertheless,
it makes still good sense to keep considering the
corresponding limiting  formal
analogue of Schr\"{o}dinger equation
 \be
 H^{(EP)}\,Q = Q\, S^{(EP)}\,.
 \label{seat}
 \ee
The columns of the so called transition matrix $Q$
can be interpreted here as forming an analogue of the
conventional eigenbasis. Also
the standard diagonal matrix of the bound-
or resonant-state eigenvalues $\widehat{E}$
is merely
replaced here by its canonical
block-diagonal analogue
 \be
 S^{(EP)}= \left[ \begin {array}{ccc}
 J^{(N_1)}(E_1)&0&\ldots\\
 \noalign{\medskip}0&J^{(N_2)}(E_2)&\ddots
 \\
 \noalign{\medskip}\vdots
 &\ddots&\ddots
 \end {array} \right]\,.
 \label{sepie}
 \ee
This matrix is, in general,
composed of individual $N_n$ by $N_n$ Jordan-block submatrices
 $$
 J^{(N_n)}(E_n)
 =\left[ \begin {array}{cccc}
 E_n&1&0&\ldots\\
 \noalign{\medskip}0&E_n&\ddots&\ddots\\
 \noalign{\medskip}\vdots&\ddots&\ddots&1\\
 \noalign{\medskip}0&\ldots&0&E_n
 \end {array}
 \right]\,,\ \ \ \ n=0,1,\ldots\,.
 $$
Every such a submatrix degenerates to the single
matrix element $J^{(1)}(E_n)=E_n$
in all of the special cases with $N_n=1$.
After a return to the strictly diagonalizable Hamiltonians, in particular,
we may simply set here
$N_1=N_2=\ldots = 1$.
For all of the non-diagonalizable, EP-admitting
Hamiltonians $H^{(EP)}$,
on the contrary,
at least
one of the dimension-denoting
superscripts
$N_k$ will be different from one in Eq.~(\ref{sepie}).
At the same time,
we will skip the models with degeneracies
$E_n=E_k$ as $n \neq k$ as
tractable (cf., e.g., \cite{without}) but,
for our present purposes, too artificial

The usefulness
of the study of the
Schr\"{o}dinger-like (albeit manifestly unphysical)
EP-related eigenvalue
problem (\ref{seat})
will be twofold.
Firstly, it will enable us to see what happens with the
conventional wave functions during the
collapse $\widehat{\Psi} \to Q$.
Secondly, for the Hamiltonians of the phenomenologically
extremely interesting EP-perturbed
form $H=H^{(EP)} + \lambda\,V$
(with any suitable bounded matrix $V$
or matrix function $V=V(\lambda)$)
one can develop some dedicated perturbation
methods of construction of the states of the system
in some small vicinity
of the EP singularity (cf., e.g., \cite{admissible,admissibleb}).

\section{Hamiltonians in the vicinity of EPs\label{sectionpert}}

Truncated matrix models with a not too small $N$
can often be treated as reasonably reliable approximate partners
of
a realistic infinite-dimensional Hamiltonian.
In such a case, given a suitable truncated but diagonalizable $N$ by $N$
matrix $H=H^{(N)}$, one can treat it as a Hamiltonian
(i.e., as the generator of evolution
of the state-vectors $|\psi(t)\kt$)
of a quantum system represented
in Schr\"{o}dinger picture.
In this sense even the choice of
$N=8$ would not already be too small, so one might expect that
even such a simplified model would already mimic
and reflect multiple innovative
spectral-design concepts and versions of some
nontrivial EP-related
phenomena. Naturally, the price to pay is that
whenever the dimension $N$ grows,
the secular determinant
becomes polynomial of a rather high degree
so that
in the general case the localization of the spectrum
would be a purely numerical task.

\subsection{The perturbed Schr\"{o}dinger equation}

The evaluation of the
predictions of the
measurements usually starts from
the solution of the time-independent
Schr\"{o}dinger equation, indeed \cite{Messiah}.
In our present considerations, therefore, we
have to
return
from the anomalous Eq.~(\ref{seat})
to
the standard Schr\"{o}dinger bound-state
eigenvalue problem
 \be
 H\,\widehat{\Psi} = \widehat{\Psi}\, \widehat{E}\,
 \label{seoff}
 \ee
where the symbol $\widehat{E}$ denotes the
diagonal matrix of the system's eigen-energies
(which have to be, in unitary systems, real)
and where the corresponding eigenvectors form
the columns of the $N$ by $N$ matrix $\widehat{\Psi}$.

Let us now be interested
in the behavior of the $N-$level quantum system in a
small vicinity of its isolated
EP singularity of any order $M$, EP=EP(M).
At the large $N \gg 1$,
the task would naturally require the
evaluation of
an overcomplicated secular polynomial
which could hardly be used
as an implicit definition of the spectrum.
Even for the general tridiagonal $N=8$ matrix, for example,
the mere
printed version of
its form provided by MAPLE \cite{Maple}
appeared to need 13 rows.
For this reason it makes sense to
turn attention to the simplified
class of Hamiltonians (\ref{umohs}) in what follows.

Multiple properties of such systems
can be described using the {\it ad hoc\,} perturbation-expansion
technique of paper \cite{admissible}. Let us now recall
several basic features of this technique.
Firstly, let us emphasize that
in \cite{admissible}
we have shown that
the above-mentioned goals cannot be reached using the standard
Rayleigh-Schr\"{o}dinger perturbation expansions.
In the innovative, EP-related scenarios
the unperturbed Hamiltonian $H^{(EP)}$ itself
is non-diagonalizable and, hence, unphysical.
In its vicinity, nevertheless, one can reveal
multiple analogies
with the conventional perturbation theory of textbooks.

In the initial technical steps of the EP-related
perturbation-approximation construction
of the solution of our
present non-degenerate
perturbation problem
(\ref{seoff}) with
 $$
  H= H^{(EP)}+ \lambda\, V(\lambda)
 $$
the analogy
with the textbooks
is in fact virtually complete,
especially because we are going to deal here with
the simplified, truncated,
finite-dimensional matrices.
This enables us to expect the knowledge
of the complete solution $Q$ of the
unperturbed Eq.~(\ref{seat}). In the subsequent step we
transform the Hamiltonian without changing its spectrum,
 \be
 H^{(EP)}+  \lambda\,V(\lambda)\ \to \
 Q^{-1}\,[H^{(EP)}+ \lambda\, V(\lambda)]\,Q
 = S^{(EP)} +  W(\lambda)\,,
 \label{pisso}
 \ee
i.e., we
rewrite our perturbed Schr\"{o}dinger equation (\ref{seoff})
using the EP-related analogue of unperturbed basis $Q$.
The linear algebraic eigenvalue problem
is then obtained,
 \be
 [S^{(EP)} +  W(\lambda)]\,\widehat{\Psi}( \lambda)
  = \widehat{\Psi}( \lambda)\, \widehat{E}( \lambda)\,.
 \label{pissoff}
 \ee
Without any methodical loss we may ignore the
Hermiticity-related
part of the construction
(i.e., all of the one-dimensional $N_k=1$ submatrix components
of $S^{(EP)}$ in (\ref{sepie}) if any)
as known and conceptually trivial.
Moreover, we will also simplify
our argumentation by assuming that
just one of the Jordan blocks is nontrivial.

This means that
after the introduction of the difference
${\epsilon}_n=E_n(W)-E^{(EP)}$
and
after such a choice (i.e., {\it ad hoc\,} shift)
of the origin of the
energy scale that the unperturbed energy itself
becomes equal to zero,
$E^{(EP)}=0$,
we have to
solve just the $N-$plet of equations
 \be
 [J^{(N)}(0) +  W( \lambda)]
 \,|\overrightarrow{\psi^{(n)}}( \lambda)\kt =
 \epsilon_n( \lambda) \,
 |\overrightarrow{\psi^{(n)}}( \lambda)\kt\,,
 \ \ \ \ n=1,2,\ldots,N
 \,.
 \label{issoff}
 \ee
In a way noticed in \cite{admissible} this implies that
the value of ${\epsilon}_n$ (with an arbitrary subscript)
can be selected as
playing the role of an alternative
measure of the smallness of the perturbation.

\subsection{Leading-order solution}

The preceding remark enables one to reconsider
the criteria of the existence
of at least one
channel of the strictly unitary evolution of the
system (see \cite{admissibleb} for details). Nevertheless,
our present task is different:
We will be interested in a classification of the
unfoldings, i.e., we will need to know {\it all\,}
of the solutions of Eq.~(\ref{issoff}), not just the
special, unitarity-compatible one.
In other words,
we will need to
know,
near a preselected EP(N) singularity, the behavior of the
whole
spectrum $\{\epsilon_n\}$
under a generic perturbation.

Naturally, the latter spectrum is, in general, complex.
For its construction one may still
follow the algebra as presented in \cite{admissible}.
For this purpose
we will abbreviate
$x=|\psi^{(n)}_1\kt$ and $|\psi^{(n)}_j\kt=y_{j-1}$ at
$j=2,3,\ldots,N$, introduce a redundant variable
$y_{N}$
and choose the
normalization $x=1$.
Moreover, we will
define the two  mutually inverse
$N$ by $N$ matrices
 \be
 L=L(\epsilon)=\left (
 \begin{array}{ccccc}
 1&0&0&\ldots&0\\
 -\epsilon&1&0&\ddots&\vdots\\
 0&-\epsilon&\ddots&\ddots&0\\
 \vdots&\ddots&\ddots&1&0\\
 0&\ldots&0&-\epsilon&1
 \ea
 \right )\,,
 \ \ \ \
 R=R(\epsilon)=\left (
 \begin{array}{ccccc}
 1&0&0&\ldots&0\\
 \epsilon&1&0&\ddots&\vdots\\
 \epsilon^2&\epsilon&\ddots&\ddots&0\\
 \vdots&\ddots&\ddots&1&0\\
 \epsilon^{N-1}&\ldots&\epsilon^2&\epsilon&1
 \ea
 \right )\,
 \label{9}
 \ee
and also the $N$ by $N$ interaction-representing matrix
 \be
 Z=\left (
 \begin{array}{cccc}
 W_{1,2}&W_{1,3}&\ldots&W_{1,N+1}\\
 W_{2,2}&W_{2,3}&\ldots&\ldots\\
 \ldots&\ldots&\ldots&\ldots\\
 W_{N,2}&W_{N,3}&\ldots&W_{N,N+1}
 \ea
 \right )\,
 \ee
where, in comparison with $W$, we omitted the first column
and added
the trivial last column with $W_{j,N+1}=0$ at
all $j$.

In terms of these matrices  we may reinterpret
our perturbed Schr\"{o}dinger  Eq.~(\ref{issoff})
as the following inhomogeneous linear algebraic problem
 \be
 (L + Z) \vec{y}= \vec{r}\,
 \label{20}
 \ee
where we abbreviated
 \be
 \vec{y} = \left (
 \ba
 y_1\\
 y_2\\
 \vdots\\
 y_{N}
 \ea
 \right )\,,
 \ \ \ \ \ \
 \vec{r} = \left (
 \ba
 \epsilon - W_{1,1}\\
 -W_{2,1}\\
 \vdots\\
 -W_{N,1}
 \ea
 \right )\,.
 \label{tarov}
 \ee
The two
alternative unperturbed equations
(\ref{issoff}) and (\ref{20})
remain equivalent,
provided only that we
accompany the latter one by the additional
mathematical equivalence requirement
 \be
 y_N=0\,.
 \label{coco}
 \ee
Now
we may eliminate the vector $\vec{y}$ out of Eq.~(\ref{20})
and expand it formally in a power series with respect to
the operator of perturbation $Z$,
 \be
 \vec{y}= R(\epsilon)\,\vec{r}-
 R(\epsilon)\,Z\,
 R(\epsilon)\,\vec{r}+R(\epsilon)\,Z\,R(\epsilon)\,Z\,
 R(\epsilon)\,\vec{r}
 - \ldots\,.
 \label{tadef}
 \ee
Whenever one decides to keep the
matrix elements of the perturbation
uniformly bounded,
 \be
 |W_{j,k}| = {\cal O}(\lambda)\,,
 \label{limes}
 \ee
we may truncate the
infinite series (\ref{tadef}) and
keep just its dominant part,
 \be
 \left (
 \ba
 y_1\\
 y_2\\
 \vdots\\
 y_{N}
 \ea
 \right )\,\approx \,
 \left (
 \begin{array}{ccccc}
 1&0&0&\ldots&0\\
 \epsilon&1&0&\ddots&\vdots\\
 \epsilon^2&\epsilon&\ddots&\ddots&0\\
 \vdots&\ddots&\ddots&1&0\\
 \epsilon^{N-1}&\ldots&\epsilon^2&\epsilon&1
 \ea
 \right )\,\left (
 \ba
 \epsilon - W_{1,1}\\
 -W_{2,1}\\
 \vdots\\
 -W_{N,1}
 \ea
 \right )\,.
 \ee
The self-consistency
constraint (\ref{coco}) then degenerates to the
explicit approximate secular equation
 \be
 (\epsilon - W_{1,1})\,\epsilon^{N-1}
 -W_{2,1}\,\epsilon^{N-2}- \ldots
 -W_{N-1,1}\,\epsilon-W_{N,1}=0\,.
 \label{firstord}
 \ee
In the immediate vicinity of the EP limit this equation solely
depends on the perturbations represented by the
first column of
matrix $W$. Thus,
only this column can control the behavior of the
energy spectrum.
Moreover, under assumption (\ref{limes})
such a leading-order secular equation further
degenerates to its asymptotically dominant form
 \be
 \epsilon^{N}
 -W_{N,1}=0\,.
 \label{firstord1}
 \ee
This observation can be re-read as the proof of the following result.

\begin{lemma}
 \label{lemma1}
Whenever $W_{N,1}\neq 0$
is a non-vanishing complex or real number or
a bounded ${\cal O}(\lambda)$ function of $\lambda$,
the unfolding of the EP(N) singularity becomes
prescribed by the leading-order formula
 \be
 {\epsilon}_n
  \approx \left [W_{N,1}(\lambda)
  \right ]^{1/N} (\lambda)\, \exp (2\pi {\rm i}n/N)
 + {\rm corrections}\,,
 \ \ \ \ n = 1,2,\ldots, N\,.
 \label{rees}
 \ee
 \end{lemma}

\begin{cor}
Whenever $W_{N,1}\neq 0$,
the rest of the perturbation $W$ can only influence the
higher-order corrections.
\end{cor}

 \noindent
One might also add that for the subclass of perturbations
characterized by the vanishing element $W_{N,1}= 0$
or by the subdominance of function $W_{N,1}(\lambda)= o(\lambda)$,
the inspection of Eq.(\ref{firstord}) reveals that
the removal of the EP degeneracy becomes incomplete.
Moreover, one cannot proceed in parallel with Lemma \ref{lemma1}
and
deduce that
at a
non-vanishing value of $W_{N-1,1}$ one gets
 $ {\epsilon}_1=0$ and
 ${\epsilon}_n
  ={\cal O} (\lambda^{1/(N-1)})\ $
for $ n = 2,3,\ldots, N$
because in such a case one would have to take into account also the
second term in the resolvent-expansion series (\ref{tadef}).

\subsection{Fine-tuned anomaly $W_{N,1}=0$ in an illustrative $N=4$ example}

%\subsection{bodovy Big Bang}fine-tuned

Although the hypothesis
 $W_{N,1}\neq 0$
of Lemma \ref{lemma1} can be considered generic,
it implies that
in the vicinity of EP(N) with $N>2$, the spectra
are not real (cf. Eq.~(\ref{rees}))
so that the
generic quantum systems are all
non-unitary.
In a way pointed out in \cite{corridors},
the requirement of unitarity would imply the necessity of
an {\it ad hoc\,} fine-tuning of the perturbation.

For illustration purposes
let us pick up the following
four-by-four toy-model Hamiltonian
of Ref.~\cite{tridiagonal},
 \be
 H^{(4)}_{\tiny \cite{tridiagonal}}(t)=\left[ \begin {array}{cccc}
  -3&\sqrt {3-3\,t}&0&0\\
  \noalign{\medskip}-\sqrt {3-3\,t}&-1&2\,\sqrt {1-t}&0\\
  \noalign{\medskip}0&-2\,\sqrt {1-
 t}&1&\sqrt {3-3\,t}\\
 \noalign{\medskip}0&0&-\sqrt {3-3\,t}&3
 \end {array} \right]\,.
 \label{foss}
 \ee
In this case the computer-assisted solution
of Eq.~(\ref{seoff}) yields
the (unordered) spectrum of
the bound-state energies $\{\sqrt{t},
-\sqrt{t},3\,\sqrt{t},-3\,\sqrt{t}\}$
which is real for $t\geq 0$ and which,
in the light of Lemma \ref{lemmaNmax}, degenerates
to the EP(4) singularity
in the limit
$t \to 0^+$.

The computer-assisted solution
of the generalized Schr\"{o}dinger Eq.~(\ref{seat}) reconfirms that
in this limit we really get the
complete EP degeneracy $E_n \to 0$.
This means that we have to choose
$N=N_1=4$ in $S^{(EP)}$
containing just the single Jordan block.
Serendipitously we
reveal that also the evaluation of
the transition matrix $Q=Q^{(N)}$
remains non-numerical in this case,
 \be
 Q^{(4)}=\left[ \begin {array}{cccc} -6&6&-3&1
 \\\noalign{\medskip}-6\,\sqrt {3}&4\,\sqrt {3}&-\sqrt {3}&0
 \\\noalign{\medskip}-6\,\sqrt {3}&2\,\sqrt
 {3}&0&0\\\noalign{\medskip}-6&0&0&0\end {array} \right]\,.
 \label{tran4}
 \ee
This means that the transformation
$H^{(4)}_{\tiny \cite{tridiagonal}}(t)
 \to Q^{-1}H^{(4)}_{\tiny \cite{tridiagonal}}(t)Q
 =J^{(4)}+W^{(4)}_{\tiny \cite{tridiagonal}}(t)$
of Eq.~(\ref{pisso})
yields the perturbation $W$ in the compact and exact form
 $$
 W^{(4)}_{\tiny \cite{tridiagonal}}(t)=
 \left[ \begin {array}{cccc} -3\,{\it \eta}&{\it \eta}&0&0
\\\noalign{\medskip}-6\,{\it \eta}&-{\it \eta}&{\it \eta}&0
\\\noalign{\medskip}0&-8\,{\it \eta}&{\it \eta}&{\it \eta}
\\\noalign{\medskip}0&0&-6\,{\it \eta}&3\,{\it \eta}\end {array} \right]
 $$
where we defined $\eta=\eta(t)=\sqrt{1-t}-1 \approx -t/2-t^2/8-\ldots$.
In this example, therefore, we obtain not only the fine-tuning of the
decisive, dominating perturbation element
$W_{N,1}=0$ but also the further ``anomalous'' fine-tuned values of
subdominant
$W_{N-1,1}=0$ and $W_{N,2}=0$.

In the next section we intend to complement these observations by
a proposal of a certain exactly tractable class of
benchmark models
of an arbitrary matrix dimension $N=2J$
in which
$W_{N,1}\neq 0$,
i.e.,
in which the perturbation $W$ would not have to be
fine-tuned.

\section{Step-by-step Hermitizations in the generic case}

For a generic perturbation with $W_{N,1}\neq 0$
and with $N>2$,
Lemma \ref{lemma1} and its corollary tell us that
in the EP(N) vicinity
the perturbed spectrum
cannot be all real.
Let us now accompany such an observation by
another illustrative example.

\subsection{Elementary generic $N=4$ model}

One of the mathematically most important features of the
fine-tuned four-by-four perturbed model (\ref{foss}) is that
in the definition (\ref{repa}) of its $t-$dependent
(i.e., perhaps, time-dependent) off-diagonal matrix elements $a_j=a_j(t)$
the EP(4)-unfolding process is controlled by the fine-tuned
and, hence, manifestly
$j-$dependent coefficients in $z_j(t)= z_j^{(EP)}-c_j\,t$ with
$c_1=3$ and $c_2=4$.
Without such a fine-tuning ,
i.e., say, after a ``more natural'' $j-$independent redefinition of
$c_1=c_2=1$
one would get an alternative toy-model
Hamiltonian
 $$
 H^{(4)}(t)=\left[ \begin {array}{cccc} -3&\sqrt {3-t}&0&0\\\noalign{\medskip}-
\sqrt {3-t}&-1&\sqrt {4-t}&0\\\noalign{\medskip}0&-\sqrt {4-t}&1&
\sqrt {3-t}\\\noalign{\medskip}0&0&-\sqrt {3-t}&3\end {array} \right]\,.
 $$
%
%nas J+W
%
%% $$
% \left[ \begin {array}{cccc} 3-\sqrt {3}\sqrt {3-t}&1/3\,
%\sqrt {3}\sqrt {3-t}&0&0\\\noalign{\medskip}6-4\,\sqrt {3}\sqrt {3-t}+3\,\sqrt
%{4-t}&-2\,\sqrt {4-t}+1+\sqrt {3}\sqrt {3-t}&1/2\,\sqrt {4-t}&0
%\\\noalign{\medskip}-12\,\sqrt {3}\sqrt {3-t}+18\,\sqrt {4-t}&4\,
%\sqrt {3}\sqrt {3-t}+8-10\,\sqrt {4-t}&-\sqrt {3}\sqrt {3-t}-1+2\,
%\sqrt {4-t}&1/3\,\sqrt {3}\sqrt {3-t}\\\noalign{\medskip}-24\,\sqrt {3
%}\sqrt {3-t}+36\,\sqrt {4-t}&12\,\sqrt {3}\sqrt {3-t}-18\,\sqrt {4-t}&
%6-4\,\sqrt {3}\sqrt {3-t}+3\,\sqrt {4-t}&-3+\sqrt {3}\sqrt {3-t}
%\end {array} \right]
% $$
%
At the sufficiently large values of $t>4$ this matrix becomes Hermitian
so that the spectrum will be all real and discrete.
In the innocent-looking limit $t\to 4^+$ the matrix
will degenerate to the direct sum of the two
decoupled two-by-two Hermiitan submatrices,
but below this value of $t$ (or, more precisely,
inside the interval of $t \in (3,4)$)
we obtain the first nontrivial example of the Hermitian--non-Hermitian
interaction. It becomes proportional to the coupling strength $\sqrt{3-t}$
and it will
connect the non-Hermitian central two-by-two submatrix
(or, more precisely, the submatrix $C$ of $H_2$ in Eq.~(\ref{speci}))
with the diagonal (i.e., Hermitian) matrix $H_1$.

Naturally, the spectrum of $H^{(4)}$ and, more specifically, the
nontrivial part of the spectrum of $H_2$
(i.e., the spectrum of $C$) ceases to be real below the EP(2)
limit of $t\to 3^+$.
At the same time, at $t<3$ the Hermitian
component disappears
and the spectrum will have the two real
components down to the ultimate EP(4)
limit of $t\to 0^+$.

In the above-described perturbative small$-t$
dynamical regime the analysis of the role of perturbations
may proceed
along the same lines as above. Transformation
$$H^{(4)}(t) \to Q^{-1}H^{(4)}(t)Q=J^{(4)}+W^{(4)}(t)$$
%of Eq.~(\ref{pisso})
will yield the compact and exact perturbation matrix
 $$
 W^{(4)}= \left[ \begin {array}{cccc} {\it \eta_1}&-{\it \eta_1}/3&0&0
 \\\noalign{\medskip}4\,{\it \eta_1}-3\,{\it \eta_2}/2&{\it \eta_2}-{\it \eta_1}&-
 {\it \eta_2}/4&0\\\noalign{\medskip}12\,{\it \eta_1}-9\,{\it \eta_2}&-4\,{
\it \eta_1}+5\,{\it \eta_2}&{\it \eta_1}-{\it \eta_2}&-{\it \eta_1}/3
\\\noalign{\medskip}24\,{\it \eta_1}-18\,{\it \eta_2}&-12\,{\it \eta_1}+9\,{
\it \eta_2}&4\,{\it \eta_1}-3\,{\it \eta_2}/2&-{\it \eta_1}\end {array} \right]
 $$
where we just abbreviated
$$\eta_1=\eta_1(t)=3-3\,\sqrt{1-t/3} \approx t/2+t^2/24 +{\cal O}(t^3)$$ and
$$\eta_2=\eta_2(t)=4-4\,\sqrt{1-t/4} \approx t/2+t^2/32 +{\cal O}(t^3).$$
Thus, it is possible to
conclude that at small $t$ one has $W_{4,1}= 3\,t+7\,t^2/16+{\cal O}(t^3)$.
This means that
in the light of the leading-order perturbation formula (\ref{rees})
there exist no real energy roots at the
small and negative times $t<0$, while there emerge
strictly two real roots
(plus the two purely imaginary roots) at
the small and positive $t>0$.
%
%nase substituce in W (kladne eps = asi t/2 v dominante)
%
%%fur1:=subs({sqrt(3-t)=sqrt(3)*(1-\eta_1/3),sqrt(4-t)=2-\eta_2/2},fu1);
%
%takze tau je rovno 21 t a EP(4) je komplexni krizOne-parametric l

%
%\begin{table}[h]
%\caption{Dimensions in partitioning
%(\ref{speci})
%under assumption (\ref{ordered}).}
% \label{Pwe1}
%\centering
%\begin{tabular}{||c||c|c|c|c|c|c||}
%\hline \hline
% &${z}_1>0$
% &${z}_1\leq 0$
% &${z}_2\leq 0$
% &\ldots
% &${z}_{J-1}\leq 0$
% &$-$
% %&${z}_J<0$
% \\
% &$-$
% &${z}_2> 0$
% &${z}_3> 0$
% &\ldots
% &${z}_{J}> 0$
% &${z}_J= 0$
% %&${z}_J<0$
% \\
% \hline \hline
% $M$
% &$-$
% &$1$
% &$2$
% &\ldots
% &$J-1$
% &$J$
% %&-
% \\
% \hline \hline
% $K$
% &$J$
% &$J-1$
% &$J-2$
% &\ldots
% &$1$
% &$-$
% %&-
% \\
% \hline \hline
%\end{tabular}
%\end{table}
%as follows, (cf. Table~\ref{Pwe1})
% \be
% {z}_1 < {z}_2 < \ldots < {z}_J \,.
% \label{ordered}
% \ee

\subsection{Linear unfoldings at arbitrary $N=2J$}

Encouraged by our preceding example
we will postulate, at any dimension $N=2J$,
 \be
 a_j(t)=
 \sqrt{z_j(t)}=
 \sqrt{z_j^{(EP(N))}-t}\,,\ \ \ \
 j = 1, 2, \ldots, J\,,\ \ \ \ t \in \mathbb{R}\,.
 \label{undered}
 \ee
At any real shift {\it alias\,} ``time'' $t$
these values of the off-diagonal matrix elements
of our toy-model
Hamiltonians (\ref{umohs}) will be ordered
according to Eq.~(\ref{ordered}).
In the light of Lemma \ref{lemmaNmax}, naturally,
such an ordering
is guaranteed at any $t$.

After partitioning,
the latter ordering convention
enables us to control the
Hermitian and non-Hermitian
sub-matrix dimensions $M$ and $2K$ in Eqs.~(\ref{quando}) and (\ref{speci}).
Indeed, the changes of these dimensions can only occur
at the instants when the Hermitian -- non-Hermitian coupling disappears,
i.e., when $U_\pm=V_\pm=0$, i.e., when we pick up
one of subscripts $j_0 = 1, 2, \ldots, J$ and choose
$t=z_{j_0}^{(EP(N))}$.

At the latter parameter $t=t_{j_0}^{(EP(N))}$
the respective function $z_{j_0}=z_{j_0}(t)$ passes through its simple zero and
changes its sign. Hence, the growth of $t$ implies the growth of $M \to M+1$
and the decrease of $K \to K-1$.
In this context it is useful to make the following
elementary observation.
\begin{lemma}
Functions $f(j,N)=z_j^{(EP(N))}=j\,(N-j)\,$ entering Eq.~(\ref{undered})
satisfy the recurrence relation
 $$
 f(j,N)=f(j-1,N-2)+f(1,N-1)
 \,
 $$
at any $j$ and $N$.
 \label{lemmatriv}
\end{lemma}
As a consequence,
any ``initial'' $t-$dependent Hamiltonian
matrix $H^{(N)}=H^{(N)}(t)$ of Eq.~(\ref{umohs})
with the elements $a_j$ of Eq.~(\ref{repa}) which are
defined in terms of the
linearly
$t-$dependent parameters $z_j$ in Eq.~(\ref{undered})
can be interpreted as the
partitioned matrix $H^{(N)}$ of Eq.~(\ref{speci}) with
the
$t-$dependent Hermitian-submatrix and non-Hermitian-submatrix dimensions
$M=M(t)$ and $2K=2K(t)$, respectively.
With the growth of $t$ the growth of $M(t)$ (by one) is encountered
when the coupling passes through zero.
In Eq.~(\ref{speci}), in the light of Lemma \ref{lemmatriv},
the coupling disappears precisely at an EP(2K)
value of the parameter so that we have
 \be
 H^{(N)}(t)=
 \left[ \begin {array}{c|c|c}
 A&0&0\\
 \hline
 0&C&0
 \\
 \hline
 0&0&B
 \end {array} \right]\,,
 \ \ \ \ \
 t=t^{(EP(2K))}\,,
 \ \ \ \ \
 A=A^\dagger\,,
 \ \ \ \ \
 B=B^\dagger\,,
 \ \ \ \ \
 C\neq C^\dagger
 \,.
 \label{uspeci}
 \ee
The details of such a coincidence will be discussed in the next
subsection.

\subsection{Illustrative model with $N=8$}

The choice of $N=2J=8$ is already
sufficiently nontrivial and suitable for illustration purposes.
With the $J-$plet of the
linear functions (\ref{undered}) of $t$ at $J=4$, viz., with
 \be
 z_1(t)={7}-t;
 \ \ \ \
 z_2(t)={12}-t;
 \ \ \ \
 z_3(t)={15}-t;
 \ \ \ \
 z_4(t)={16}-t
 \label{urdo}
  \ee
we already have the possibility of the
application of Lemma \ref{lemmatriv}
because when we set $t=z_1(0)=7$ we get
the new set of tilded parameters
 \be
 \widetilde{z_1}(t)=0;
 \ \ \ \
 \widetilde{z_2}(t)=5;
 \ \ \ \
 \widetilde{z_3}(t)=8;
 \ \ \ \
 \widetilde{z_4}(t)=9
 \label{urdo7}
  \ee
at which the original
EP(8)-supporting $t=0$ Hamiltonian (cf. Eq.~(\ref{onyctyri}))
becomes replaced by the $t=7$ Hamiltonian
matrix
$$
H^{(8)}(t=7)=
 \left[ \begin {array}{cccccccc} -7&0&0&0&0&0&0&0\\\noalign{\medskip}0
&-5&\sqrt {5}&0&0&0&0&0\\\noalign{\medskip}0&-\sqrt {5}&-3&\sqrt {8}&0
&0&0&0\\\noalign{\medskip}0&0&-\sqrt {8}&-1&\sqrt {9}&0&0&0
\\\noalign{\medskip}0&0&0&-\sqrt {9}&1&\sqrt {8}&0&0
\\\noalign{\medskip}0&0&0&0&-\sqrt {8}&3&\sqrt {5}&0
\\\noalign{\medskip}0&0&0&0&0&-\sqrt {5}&5&0\\\noalign{\medskip}0&0&0&0
&0&0&0&7\end {array} \right]\,.
$$
The
non-Hermitian component of the
latter matrix is precisely
the $K=6$ submatrix $C$ of the decoupled
model~(\ref{uspeci}) where we have
$N=8$ and $M=1$, i.e., where the
``one-by-one Hermitian submatrices'' $A$ and $B$ are trivial
(i.e., ``diagonal'', equal to the mere
numbers $-7$ and $7$, respectively).
The central non-Hermitian six by six submatrix $C$ itself
is, in the light of Lemma \ref{lemmaNmax}, non-diagonalizable and
exhibiting the EP(6) degeneracy.

At $t=7$,  the perturbation-approximation analysis
of section \ref{sectionpert}
is directly applicable. The
necessary transition-matrix solution $Q^{(8)}$ of the corresponding
generalized Schr\"{o}dinger Eq.~(\ref{seat})
becomes available in the following closed form at $2K=6$,
 $$
 Q^{(8)}(t=7)=
 \left[ \begin {array}{cccccccc} 0&0&0&0&0&0&1&0\\\noalign{\medskip}-
120&120&-60&20&-5&1&0&0\\\noalign{\medskip}-120\,\sqrt {5}&96\,\sqrt {
5}&-36\,\sqrt {5}&8\,\sqrt {5}&-\sqrt {5}&0&0&0\\\noalign{\medskip}-60
\,\sqrt {5}\sqrt {8}&36\,\sqrt {5}\sqrt {8}&-9\,\sqrt {5}\sqrt {8}&
\sqrt {5}\sqrt {8}&0&0&0&0\\\noalign{\medskip}-20\,\sqrt {5}\sqrt {8}
\sqrt {9}&8\,\sqrt {5}\sqrt {8}\sqrt {9}&-\sqrt {5}\sqrt {8}\sqrt {9}&0
&0&0&0&0\\\noalign{\medskip}-40\,\sqrt {5}\sqrt {9}&8\,\sqrt {5}\sqrt
{9}&0&0&0&0&0&0\\\noalign{\medskip}-40\,\sqrt {9}&0&0&0&0&0&0&0
\\\noalign{\medskip}0&0&0&0&0&0&0&1\end {array} \right]\,.
 $$
After a trivial rearrangement of the basis we can see
that as expected, this is an eight by eight matrix
which is equal to a direct sum of
a non-trivial six by six transition matrix
(related to the non-diagonalizable submatrix $C$
of the $t=7$ EP(6)-supporting Hamiltonian) with the
other two decoupled diagonal real elements
normalized to one.

In the next step we have to choose
$t=12$
and get the other Hamiltonian
matrix
$$
H^{(8)}(t=12)=
\left[ \begin {array}{cccccccc} -7&{\rm i}\,\sqrt {5}&0&0&0&0&0&0
\\
\noalign{\medskip}-{\rm i}\,\sqrt {5}&-5&0&0&0&0&0&0\\\noalign{\medskip}0&0&
-3&\sqrt {3}&0&0&0&0\\
\noalign{\medskip}0&0&-\sqrt {3}&-1&\sqrt {4}&0&0
&0\\\noalign{\medskip}0&0&0&-\sqrt {4}&1&\sqrt {3}&0&0
\\
\noalign{\medskip}0&0&0&0&-\sqrt {3}&3&0&0\\\noalign{\medskip}0&0&0&0
&0&0&5&{\rm i}\,\sqrt {5}\\\noalign{\medskip}0&0&0&0&0&0&-{\rm i}\,\sqrt {5}&7
\end {array} \right]\,.
$$
This example offers the first nontrivial
realization of the block-diagonal pattern (\ref{uspeci})
with $M=2$ and with the four-by-four
EP(4)-supporting submatrix $C=C^{(EP(4))}$.

In the latter example, a minor technical comment is due,
emphasizing that the treatment of the non-Hermitian,
EP(4)-related  four by four matrix part of the
Hamiltonian remained easy. Also the construction of the related
transition
matrix $ Q^{(8)}(t=12)$ remained straightforward
yielding the compact formula for the corresponding decoupled
submatrix
 $$
 Q^{(4)}=
 \left[ \begin {array}{cccc}
-6&6&-3&1\\\noalign{\medskip}-6\,\sqrt {3}&4\,\sqrt {3}&-
\sqrt {3}&0\\\noalign{\medskip}-3\,\sqrt {3}\sqrt {4}&\sqrt {3
}\sqrt {4}&0&0\\\noalign{\medskip}-3\,\sqrt {4}&0&0&0
\end {array} \right]\,.
 $$
In contrast, the emergence of the other two
Hermitian but non-trivial two-by-two
submatrices $A$ and $B$ led to
unexpected problems. In particular,
the
standard commercial software (viz., MAPLE \cite{Maple})
produced its output
in a rather puzzling (though still correct)
form. For example,
in place of
the four elementary eigenvalues
of the two-by-two
submatrices $A$ and $B$ (viz., in place of the
four numbers $\pm 6 \pm \sqrt{6}$)
the computer-assisted computations resulted in formula
$$  6\,{\frac {-2899+1189\,\sqrt {6}}{ \left( 7\,\sqrt {6}-12
 \right)  \left( -73+28\,\sqrt {6} \right) }}\approx -3.550510257$$
in place of the equivalent but much shorter expression
$-6+\sqrt{6}\approx -3.550510257$. Similarly we received
 $$
 -30\,{\frac {\sqrt {6}-1}{7\,\sqrt {6}-12}}
\approx -8.449489743$$
in place of $-6-\sqrt{6}$, and
$$ -5\,{\frac {\sqrt {6}
\left( 7\,\sqrt {6}-12 \right) }{-54+19\,\sqrt {6}}}
\,,\ \ \ \
6\,{\frac {-73+28\,\sqrt {6}}{-54+19\,\sqrt {6}}}  $$
in place of
$6+\sqrt{6}$,
and
 $6-\sqrt{6}$, respectively. Thus, in all of these cases
we were forced to
use the ``factor'' MAPLE function
for an additional simplification of the formulae.

%
%With the 4 x 4 submatrix of the transition matrix known,
%
% $$
% \left[ \begin {array}{cccccccc}
% 0&0&0&0&-337500\,{\frac {-73+28\,\sqrt {6}}{ \left( 6+\sqrt {6} \right)
%   \left( -6+\sqrt {6} \right) ^{
%4} \left( 73\,\sqrt {6}+168 \right)  \left( -168+73\,\sqrt {6}
% \right) }}&1875\,{\frac {7+2\,\sqrt {6}}{ \left( 6+\sqrt {6} \right)
% \left( 73\,\sqrt {6}+168 \right)  \left( -168+73\,\sqrt {6} \right) }
%}&0&0\\\noalign{\medskip}0&0&0&0&-337500\,{\frac {{\rm i}\,\sqrt {5} \left( -
%19+9\,\sqrt {6} \right) }{ \left( 6+\sqrt {6} \right)  \left( -6+
%\sqrt {6} \right) ^{4} \left( 73\,\sqrt {6}+168 \right)  \left( -168+
%73\,\sqrt {6} \right) }}&1875\,{\frac {{\rm i}\,\sqrt {5} \left( \sqrt {6}+1
% \right) }{ \left( 6+\sqrt {6} \right)  \left( 73\,\sqrt {6}+168
% \right)  \left( -168+73\,\sqrt {6} \right) }}&0&0\\\noalign{\medskip}
%-6&6&-3&1&0&0&0&0\\\noalign{\medskip}-6\,\sqrt {3}&4\,\sqrt {3}&-
%\sqrt {3}&0&0&0&0&0\\\noalign{\medskip}-3\,\sqrt {3}\sqrt {4}&\sqrt {3
%}\sqrt {4}&0&0&0&0&0&0\\\noalign{\medskip}-3\,\sqrt {4}&0&0&0&0&0&0&0
%\\
%\noalign{\medskip}0&0&0&0&0&0&56250\,{\frac {\sqrt {6}}{ \left( -168
%+73\,\sqrt {6} \right)  \left( 6+\sqrt {6} \right) ^{5}}}&5/2\,{\frac
%{-19+9\,\sqrt {6}}{-168+73\,\sqrt {6}}}\\\noalign{\medskip}0&0&0&0&0&0
%&-11250\,{\frac {{\rm i}\,\sqrt {5}}{ \left( 6+\sqrt {6} \right) ^{4} \left( -
%168+73\,\sqrt {6} \right) }}&-1/2\,{\frac {{\rm i}\,\sqrt {5} \left( -73+28\,
%\sqrt {6} \right) }{-168+73\,\sqrt {6}}}\end {array} \right]
% $$
%
%and with the compact form of the remaining matrix elements.
%
%

The continuation of the tendency
of having difficulties with the
Hermitian rather than non-Hermitian components of the Hamiltonian
has been noticed also at $t=15$.
In this case the presence of the
two decoupled Hermitian components $A$ and $B$ in the Hamiltonian
$$
H^{(8)}(t=15)=
\left[ \begin {array}{cccccccc} -7&{\rm i}\,\sqrt {8}&0&0&0&0&0&0
\\\noalign{\medskip}-{\rm i}\,\sqrt {8}&-5&{\rm i}\,\sqrt {3}&0&0&0&0&0
\\\noalign{\medskip}0&-{\rm i}\,\sqrt {3}&-3&0&0&0&0&0\\\noalign{\medskip}0&0&0
&-1&1&0&0&0\\\noalign{\medskip}0&0&0&-1&1&0&0&0\\\noalign{\medskip}0&0
&0&0&0&3&{\rm i}\,\sqrt {3}&0\\\noalign{\medskip}0&0&0&0&0&-{\rm i}\,\sqrt {3}&5&
{\rm i}\,\sqrt {8}\\\noalign{\medskip}0&0&0&0&0&0&-{\rm i}\,\sqrt {8}&7\end {array}
 \right]
$$
already forced us to search for the
six related non-zero (i.e., non-EP(2)))
eigenvalues
 $$
 \mp 9.171029786\,,\ \ \
 \mp 4.311583134\,,\ \ \
 \mp 1.517387080
 $$
(as well as for the related eigenvectors)
using the universal brute-force numerical methods.

\begin{table}[h]
\caption{The list of the EP(2K) degeneracies
for the decoupled models~(\ref{uspeci})
at $N=8$.}
 \label{Pwe2}
\centering
\begin{tabular}{||c||c|c|c|c|c||}
 \hline \hline
 $t$
 &$0$
 &$7$
 &$12$
 &$15$
 &$16$
 \\
 \hline \hline
 $M=M(t)$
 &$-$
 &$1$
 &$2$
 &$3$
 &$4$
 \\
 \hline \hline
 EP(2K)
 &EP(8)
 &EP(6)
 &EP(4)
 &EP(2)
 &$-$
  \\
 \hline \hline
\end{tabular}
\end{table}

All of the degenerate EP forms of
$H^{(8)}(t)$ with $t=t^{(EP(2K))}$
may be found listed in Table \ref{Pwe2}.
Although this Table only mentions the
size $M=M(t)$ of the two Hermitian submatrices
at the EP(2K),
one can extend the assignment by
recalling the
coverage
of the neighboring $t\neq t^{(EP(2K))}$
Schr\"{o}dinger equations
by the perturbation techniques
of section \ref{sectionpert}.
The resulting
leading-order description
of the model will hold
in the
four distinct
EP(2K) vicinities
and offer a reliable picture of
the subsequent unfoldings of the
separate spectral degeneracies.

Far from the singular EP limits, i.e.,
in the middle of
the separate open and disjunct non-EP intervals of
$t \in (-\infty,0)$,
$t \in (0,7)$,
$t \in (7,12)$,
$t \in (12,15)$ and
$t \in (15,\infty)$,
the octuplet of the separate energy levels may be expected to be
represented by the real or complex smooth functions of $t$.
Numerically, such an expectation
is strongly supported by Figure \ref{ureone}
in which we displayed all of the real eigenvalues
of $H^{(8)}(t)$ as the functions of $t$ in the interval
ranging from the small negative values
up to the $t>16$ domain of full Hermiticity.
In the picture one only has to notice that
in the interval of $t \in (15,16)$
characterized by the full reality of the spectrum
the Hamiltonian itself still remains non-Hermitian.
In the language of physicists \cite{Carl}
one could speak there about the dynamical regime with
unbroken
${\cal PT}-$symmetry.

%********** Figure 1 zde
\begin{figure}[h]                     %instead of \begin{figure}[t]
\begin{center}                         %instead of \begin{center}
\epsfig{file=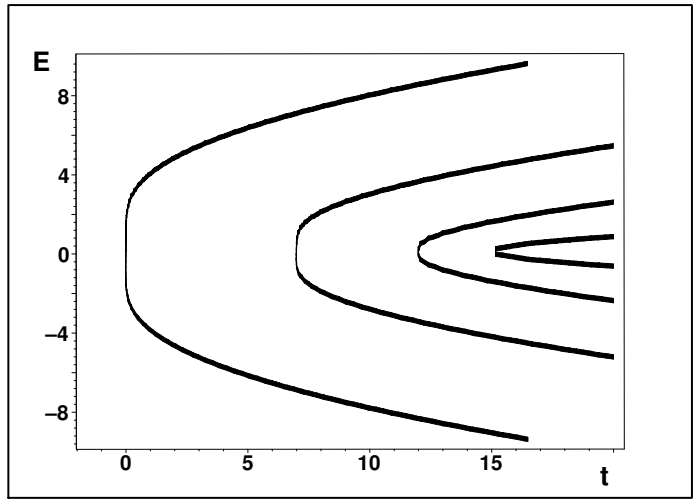,angle=270,width=0.52\textwidth}
\end{center}                         %instead of \end{center}
\vspace{-2mm} \caption{The numerically calculated
unfoldings and smoothness of the real eigenvalues of our
toy model $H^{(8)}(t)$.
 \label{ureone}}
\end{figure}

\section{Summary}

In our paper we felt motivated by the
recent enormous growth of intensity of the study
of the quantum dynamical systems living near
the EP singularities \cite{Christodoulides}.
These studies are currently opening a comparatively new domain
of quantum physics in which the challenging nature
of the experiments is paralleled by the equally challenging
nature of the necessary amendments of the conventional
mathematical tools in functional analysis \cite{SKbook}.

In the latter context it is necessary to emphasize that
the number of the newly emerging technical as well as
conceptual difficulties is enormous \cite{book}.
Thus, within such a fairly broad area of research we
decided to restrict our attention to the
questions asked by Jones \cite{Joneskonf,Jonesb}.
These questions
concern the interference between the
Hermitian and non-Heermitian components of
quantum Hamiltonians.

The difficulty of the general answers
reflects one of the
most influential currently existing terminological ambiguities,
namely, the tendency of
treating the closed quantum systems and open quantum systems
more or less on the same footing.
One of the consequences is that
the mathematically correct and rigorous theorems
may be accompanied by their
misleading phenomenological applications.
As a typical illustrative example let us recall
the concept of pseudospectrum
\cite{Trefethen}.
Naturally, such a concept
is perfectly suitable, say, for the analysis
of stability in the open quantum systems
which are characterized by the plainly
non-Hermitian Hamiltonians $H \neq H^\dagger$
with the complex spectra of resonances \cite{Nimrod}.
At the same time,
the description of the properties of the pseudospectra
is much more difficult
for the closed systems.
Indeed, as long as the evolution of the latter systems
is controlled by the
quasi-Hermitian Hamiltonians $H \neq H^\dagger$
(i.e., one must re-construct
also the correct physical inner
product metric $\Theta$ \cite{ali})
even
the specialists in the full-fledged functional analysis
find the task prohibitively difficult
\cite{PSPC}.

For this reason,
the scope of our present study of the
EP-related phenomena
was restricted to the ``toy-model'' quantum
systems for which the dimension $N$ of all of
the relevant Hilbert spaces remains finite.
The price to pay was that we had to leave
many currently studied questions
(often motivated by the urgent methodical
needs of the relativistic
quantum field theory
\cite{Carl,Carlbook,BJones}) simply unnoticed.
At the same time, a decisive benefit
of the choice of $N < \infty$
is that it is very easy to distinguish between the
models of the closed quantum systems
(in which the spectra are
easily tested to be real) and their
open-system generalizations, for which
the EP singularities
are not only experimentally detectable
but also
mathematically richer.

In the theoretical part of our paper
we had to recall
some of the known results in
the EP-related perturbation theory.
We re-emphasized that
the basic mathematical
difference emerges again
between the open- and closed-system dynamical regimes.
This observation enabled us to explain why, in the
unitary,
closed-system scenario
requiring real spectrum,
the
construction of the
admissible unfolding path in the space of parameters
near the EP singularity must be fine-tuned.
In the generic,
open-system physical scenarios, on the contrary,
it makes sense to study
all of the complex forms of the spectra.

In the constructive part of our paper
the observation of the genericity of the
complex EP(N)-unfolding patterns
was supported by
a remarkable one-parametric family of illustrative,
EP-possessing ``solvable''
toy-model Hamiltonians $H^{(N)}(t)$.
The model appeared, at any finite
Hilbert-space dimension $N=2J$,
{\em exact\,} and non-numerical,
possessing a sequence of the EPs
characterized by a decreasing degree of degeneracy
corresponding to the equally quickly decreasing dimension
of the non-Hermitian submatrix component $C$
of the Hamiltonian.

%\newpage

\section*{Appendix A: Quantum mechanics using non-Hermitian
operators}

The recent growth of popularity of the
use of various non-self-adjoint operators $\Lambda \neq \Lambda^\dagger$
in quantum physics
(see, e.g.,
several
mathematically oriented
compact recent reviews in \cite{book}) can be perceived
as motivated by the novelty of
the underlying combination of two different
domains of research in mathematics
and in physics.
In mathematics, indeed, the field is closely related to
the traditional study of
operators which are self-adjoint with respect to
an indefinite inner-product metric in
the so called Krein space ${\cal K}$ \cite{AKbook}.
In physics, in contrast,
multiple phenomenologically relevant
operators $\Lambda $ which would be self-adjoint in ${\cal K}$
are reinterpreted, in a suitable
auxiliary (i.e., manifestly unphysical
but user-friendly) Hilbert space ${\cal H}_{math}$, as
manifestly non-Hermitian but
${\cal PT}-$symmetric and, in the closed-system subcase,
${\cal PCT}-$symmetric
(here, the new symbol ${\cal C}$ denotes a charge --
see \cite{Carl}  and/or
the most
recent monographs \cite{Carlbook,Christodoulides} for explanation
and/or for the compact reviews of the current state of art).

For methodical reasons the attention is often restricted
to the most
elementary
quantum systems and models in which one takes into consideration just
a single observable, i.e., mostly, an operator $H$
representing the Hamiltonian
in Schr\"{o}dinger picture \cite{Messiah}.
Due to the
underlying, methodically innovative assumption of the
non-Hermiticity of $H$ in ${\cal H}_{math}$,
one usually
treats the theory as open-system-related and
effective,
``accepting
that this
\ldots may well involve the loss of unitarity \ldots
essentially because we are dealing with
a subsystem of a larger system whose physics has not been taken fully into
account'' \cite{Jonesb}.

In the minority of systems
one is able to reinterpret the
non-Hermiticity of $H$ in ${\cal H}_{math}$
as compatible with the Hermiticity of the same operator
in another Hilbert space (see \cite{Geyer}).
In other words, having an
operator $\Theta$ at our disposal,
one can certainly speak about a closed quantum system.
As long as the corresponding
physical, Hermitizing inner-product metric $\Theta(H)$
is at least partially non-unique, we may say that
the process of the building of
the special unitary models may rely upon
the variability of the metric
representing a certain new
degree of freedom in the theory.

Jones \cite{Jones} and
Bender with Jones \cite{BJones}
imagined that
in both the closed- and open-system contexts,
it may make sense to construct the theories
in which one admits a suitable
``coupling of non-Hermitian ${\cal PT}-$symmetric Hamiltonians to standard
Hermitian Hamiltonians'' \cite{BJones}.
We can only repeat that
such an innovation of the quantum model-building strategy
proved rewarded by the emergent possibility of control
of the quantum-phase-transition
instant marked by the loss of the unitarity
when ``the coupling becomes stronger
than some critical value'' \cite{BJones}.

In our present paper we felt challenged by the fact that the latter
results were merely
``established up to second order in
perturbation theory'' \cite{BJones}.
We developed, therefore, a
family of the exactly solvable realizations of the Hamiltonian
in which
the mathematical analysis
of the coupling between the Hermitian and non-Hermitian
components of $H$
remained transparent,
exact, non-perturbative and also basically non-numerical.

Less attention has been paid here to the
study of the conditions of the unitarity of the evolution
near EPs.
The reason is that it would be
necessary to postulate, in addition, that the
Hamiltonian $H$
appears not only non-Hermitian but also
quasi-Hermitian
in ${\cal H}_{math}$. This would mean that
this operator had to
have a strictly real spectrum. Indeed,
in the models studied in our present paper we saw that
such a condition
might happen to be over-restrictive in general.


\newpage

\begin{thebibliography}{00}



\bibitem{Messiah}
%A. Messiah, Quantum Mechanics I. Amsterdam, North Holland, 1961.
Messiah, A. \emph{Quantum Mechanics};
North Holland: Amsterdam, The Netherlands, 1961.

\bibitem{Stone}
%M. H. Stone,
%%M. H., "On one-parameter unitary groups in
%%Hilbert Space",
%Ann. Math. 33,  643 (1932).
%%–648,
%%doi:10.2307/1968538, JSTOR 1968538  (3):
Stone, M.H. On one-parameter unitary groups in Hilbert Space.
\emph{Ann. Math.} {\bf 1932}, {\em 33}, 643--648.


%\bibitem{Dieudonne}
%%J. Dieudonn\'{e},
%%% J. Quasi-Hermitian operators. In:
%%in Proc. Int. Symp. Lin. Spaces. Oxford, Pergamon, 1961, p. 115.
%%%–122.
%Dieudonne J 1961
%%. Quasi-Hermitian Operators. In
%%\emph{
%Proc Int Symp Lin Spaces
%%Pergamon, Oxford, UK,
%%{pp. 115--122}.
%%%
%%\bibitem{Dieudonne}
%%J. Dieudonne,
%%Proc. Int. Symp. Lin. Spaces
%(Pergamon, Oxford, 1961), pp. 115 - 122
%%
\bibitem{Dieudonne}
%J. Dieudonn\'{e},
%% J. Quasi-Hermitian operators. In:
%in Proc. Int. Symp. Lin. Spaces. Oxford, Pergamon, 1961, p. 115.
%%–122.
Dieudonne, J. Quasi-Hermitian Operators. In
\emph{Proc. Int. Symp. Lin. Spaces}, Pergamon: Oxford, UK, 1961,
{pp. 115--122}.


\bibitem{Geyer}
%F. G. Scholtz, H. B. Geyer and F. J. W. Hahne, Ann. Phys. (NY) 213,
%74 (1992).
%
Scholtz, F. G.;  Geyer, H. B.;
Hahne,  F. J. W.
%Quasi-Hermitian operators in Quantum Mechanics
% and the variational principle. \emph{Ann. Phys.} \textbf{1992}, \emph{213},
%74--101.Scholtz, F.G.; Geyer, H.B.; Hahne, F.J.W.
Quasi-Hermitian Operators in Quantum Mechanics and the Variational Principle.
\emph{Ann. Phys. (NY)} {\bf 1992}, {\em 213}, 74--101.
%
%\bibitem{Geyer}
%Scholtz F G, Geyer H B and Hahne F J W 1992 Ann Phys 213 74 - 101


\bibitem{Carl}
%C. M. Bender,
%%"Making Sense of Non-Hermitian Hamiltonians"
%Rep. Prog. Phys. 70, 947 (2007).
%%-1018.
Bender, C. M. Making sense of non-Hermitian Hamiltonians. \emph{Rep.
Prog. Phys.} {\bf 2007}, {\em 70}, 947--1118.
% (hep-th/0703096).
%\bibitem{Carl}
%%C. M. Bender,
%%%"Making Sense of Non-Hermitian Hamiltonians"
%%Rep. Prog. Phys. 70, 947 (2007).
%%%-1018.
%Bender C M 2007
%%Making sense of non-Hermitian Hamiltonians. \emph{
%Rep
%Prog Phys 70 947 - 1118
%% (hep-th/0703096).



\bibitem{ali}
%A. Mostafazadeh,
%%Pseudo-Hermitian Quantum Mechanics,
%%%
%%arXiv:0810.5643,
%%.
%% arXiv:0810.5643 [pdf, ps, other]
%%Ali Mostafazadeh Comments: 76 pages, 2 figures, 243 references,
%%revised version
%Int. J. Geom. Meth. Mod. Phys. 7, 1191
%%-1306
% (2010).
Mostafazadeh, A. Pseudo-Hermitian Representation of
Quantum Mechanics.
{\emph{Int. J. Geom. Meth. Mod. Phys.}} \textbf{2010}, \emph{7}, 1191--1306.
%\bibitem{ali}
%Mostafazadeh A 2010 Int. J. Geom. Methods Mod. Phys. 07 1191 - 1306
%
%\bibitem{ali}
%%A. Mostafazadeh,
%%%Pseudo-Hermitian Quantum Mechanics,
%%%%
%%%arXiv:0810.5643,
%%%.
%%% arXiv:0810.5643 [pdf, ps, other]
%%%Ali Mostafazadeh Comments: 76 pages, 2 figures, 243 references,
%%%revised version
%%Int. J. Geom. Meth. Mod. Phys. 7, 1191
%%%-1306
%% (2010).
%Mostafazadeh, A. Pseudo-Hermitian Representation of
%Quantum Mechanics.
%{\emph{Int. J. Geom. Meth. Mod. Phys.}} \textbf{2010}, \emph{7}, 1191--1306.


\bibitem{Carlbook}
%C. M. Bender, PT Symmetry in Quantum and Classical Physics.
%%with contributions from
%%P. E. Dorey, C. Dunning, A. Fring,
%%D. W. Hook, H. F. Jones, S. Kuzhel, G. Levai, and R. Tateo
%(World Scientific, Singapore, 2018).
Bender, C.M. \emph{PT Symmetry in Quantum and Classical Physics};
World Scientific: Singapore, 2018.


\bibitem{Trefethen}
Trefethen, L.N.; Embree, M. \emph{Spectra and Pseudospectra}; Princeton
University Press: Princeton, NJ, USA, 2005.


\bibitem{book}
%F. Bagarello et al, Eds,
%``Non-Selfadjoint Operators in Quantum Physics:
%Mathematical Aspects'',  Wiley, Hoboken,
%2015.
%
%F. Bagarello, J.-P. Gazeau, F. H. Szafraniec, and M. Znojil,
%Non-Selfadjoint Operators in Quantum Physics: Mathematical Aspects.
%Wiley, Hoboken, 2015.
Bagarello, F.; Gazeau, J.-P.; Szafraniec, F.; Znojil, M. (Eds.)
\emph{Non-Selfadjoint Operators in Quantum Physics: Mathematical Aspects};
Wiley: {Hoboken, NJ, USA,} %newly added information, please confirm. OK.
2015.
% ISBN: 978--1-118-85528-7, 432 pages, July 2015, pp. xvii -
%xviii, (c) 2015 John Wiley

\bibitem{Viola}
%PREPRINT of J. Math. Phys. 56 (2015), 103513
%
%Pseudospectra in non-Hermitian quantum mechanics
%
{Krej\v{c}i\v{r}\'{\i}k, D.; Siegl, P.; Tater, M.; Viola, J.}
Pseudospectra in non-Hermitian quantum mechanics. \emph{J. Math.
Phys.} \textbf{2015}, \emph{56}, 103513.
%Available on [ AIP (published version) ].
%Preprint on arXiv:1402.1082 [math-SP]. Citations: 5
%\bibitem{Viola}
%Krej\v{c}i\v{r}\'{\i}k D, Siegl P, Tater M and   Viola J 2015
%%Pseudospectra in nonhermitian quantum mechanics.
%J Math Phys  56 103513
%
%\bibitem{Viola}
%D. Krej\v{c}i\v{r}\'{\i}k, P. Siegl, M. Tater and J. Viola,
%%Pseudospectra in non-Hermitian quantum mechanics.
%J. Math. Phys. 56 (2015) 103513.
%%Available on [ AIP (published version) ].
%%Preprint on arXiv:1402.1082 [math-SP]. Citations: 5



\bibitem{Joneskonf}
Jones, H.F.
Coupling the Hermitian and pseudo-Hermitian worlds.


http://www.staff.city.ac.uk/~fring/PT
(accessed on July 20, 2022)


(transparencies
of the conference talk on July 16, 2007,
available via the PHHQP webpage).

\bibitem{Jones}
Jones, H.F.
Scattering from localized non-Hermitian potentials.
\emph{Phys. Rev. D} \textbf{2007}, \emph{76}, 125003.
%H. F. Jones
%Phys. Rev. D 76, 125003 – Published 4 December 2007
%
%%
%%We highlight the conceptual issues that arise when one applies
%the quasi-Hermitian framework to analyze scattering from localized
%non-Hermitian potentials, in particular complex square wells or delta functions.
%When treated in the framework of conventional quantum mechanics, these potentials
%are generally considered as effective theories, in which probability
%is not conserved because of processes that have been ignored. However,
%if they are treated as fundamental theories, the Hilbert-space metric
%must be changed. In order for the newly defined probability to be conserved,
%it must differ from the standard one, even at asymptotically large distances
%from the scattering center, and the mechanism for this is the nonlocality
%of the new metric, as we show in detail in the model of a single complex delta function.
%However, properties of distant bound-state systems, which do not interact physically
%with the non-Hermitian scattering potential, should not be affected.
%We analyze a model Hamiltonian that supports this contention.
%
%arXiv:0707.3031  [pdf, ps, other]  quant-ph hep-th
%doi
%10.1103/PhysRevD.76.125003
%Conceptual Problems in Scattering from Localized non-Hermitian Potentials
%
%Authors: H. F. Jones
%
%Abstract: We highlight the conceptual issues that arise when one applies the quasi-Hermitian framework
%to analyze scattering from localized non-Hermitian potentials, in particular complex square-wells or delta-functions.
%When treated in the framework of conventional quantum mechanics, these potentials are generally considered as effective theories,
%in which probability is not conserved because of processes that have been ignored. However, if they are treated as fundamental theories,
%the Hilbert-space metric must be changed. In order for the newly-defined probability to be conserved, it must differ from the standard one,
%even at asymptotically large distances from the scattering centre,
%and the mechanism for this is the non-locality of the new metric, as we show
%in detail in the model of a single complex delta function. However, properties of distant bound-state systems, which do not interact physically
%with the non-Hermitian scattering potential, should not be affected. We analyze a model Hamiltonian that supports this contention
%Submitted 26 October, 2007; v1 submitted 20 July, 2007; originally announced July 2007.
%
%Comments: The emphasis has been changed from v1, recognizing that it makes physical sense that the wave functions of scattering states are
%fundamentally changed in the quasi-Hermitian framework. In contrast, bound states should not be significantly affected by the introduction of a distant non-Hermitian scattering potential
%
%
%We highlight the conceptual problems that arise when one applies the quasi-Hermitian framework
%to analyze  scattering from localized non-Hermitian potentials, in particular complex square-wells
%or delta-functions. When treated in the framework of conventional quantum mechanics, these potentials
%are generally considered as effective theories, in which probability is not conserved because of processes that
%have been ignored. However, if they are treated as fundamental theories, the Hilbert-space metric
%must be changed. This leads to conservation of a newly-defined probability, but the change in the metric
%is not local, meaning that the entire framework of quantum mechanics is modified, even at asymptotically
%large distances from the scattering centre.\


\bibitem{MZbook}
Znojil, M. Non-Self-Adjoint Operators in Quantum Physics: Ideas, People, and Trends.
% (New
%York:Wiley) ch 1 pp 7 - 58
In
Bagarello, F.; Gazeau, J.-P.; Szafraniec, F.; Znojil, M. (Eds.)
\emph{Non-Selfadjoint Operators in Quantum Physics: Mathematical Aspects};
Wiley: {Hoboken, NJ, USA,}
2015.


\bibitem{BJones}
Bender, C.M.; Jones, H.F. Interactions of Hermitian and non-Hermitian Hamiltonians.
%Journal of Physics a: Mathematical and Theoretical. 41.
%DOI: 10.1088/1751-8113/41/24/244006
\emph{J.
Phys. A: Math. Theor.} \textbf{2008}, \emph{41}, 244006.
%
%Interactions of Hermitian and non-Hermitian Hamiltonians
%Carl M. Bender, Hugh F. Jones
%The coupling of non-Hermitian PT-symmetric Hamiltonians
%to standard Hermitian Hamiltonians, each of which individually
%has a real energy spectrum, is explored by means of a number of soluble models.
%It is found that in all cases the energy remains real
%for small values of the coupling constant, but becomes complex if the
%coupling becomes stronger than some critical value. For a quadratic
%non-Hermitian PT-symmetric Hamiltonian coupled to an arbitrary real
%Hermitian PT-symmetric Hamiltonian, the reality of the ground-state energy
%for small enough coupling constant is established up to second order in perturbation theory.
%Comments:   9 pages, 0 figures
%Subjects:   High Energy Physics - Theory (hep-th)
%Cite as:    arXiv:0709.3605 [hep-th]
%      (or arXiv:0709.3605v1 [hep-th] for this version)
%
%https://doi.org/10.48550/arXiv.0709.3605
%Focus to learn more
%Journal reference:      J.Phys.A41:244006,2008
%Related DOI:
%https://doi.org/10.1088/1751-8113/41/24/244006
%Focus to learn more
%Submission history
%From: Carl Bender [view email]
%[v1] Sat, 22 Sep 2007 21:01:12 UTC (9 KB)
%================


\bibitem{scatt}
Znojil, M.
Scattering theory with localized non-Hermiticities.
\emph{Phys. Rev. D} \textbf{2008}, \emph{78},
025026.
% (2008) (arXiv:0805.2800v1

\bibitem{discrete}
Znojil, M.
Discrete PT-symmetric models of scattering.
\emph{J.
Phys. A: Math. Theor.} \textbf{2008}, \emph{41},
292002.
%(arXiv:0806.2019v1


\bibitem{Nimrod}
Moiseyev, N. \emph{Non-Hermitian Quantum Mechanics}; CUP: Cambridge, UK, 2011.


\bibitem{Kato}
% Kato, T.  \emph{Perturbation Theory for Linear Operators}; Springer-Verlag:
%Berlin, Germnay, 1966.
Kato, T.
\emph{Perturbation Theory for Linear Operators};
Springer: Berlin/Heidelberg, Germany,
 1966.




\bibitem{denis}
{Berry, M.V.} Physics of Nonhermitian Degeneracies.
\emph{Czechosl. J. Phys.} \textbf{2004}, \emph{54}, 1039--1047.

\bibitem{denisb}
Heiss, W.D. Exceptional points - their universal occurrence and
their physical significance. \emph{Czechosl. J. Phys.}
\textbf{2004}, \emph{54}, 1091--1100.


\bibitem{denisc}
{Klaiman, S.; G\"{u}nther, U.; Moiseyev, N.} Visualization of Branch
Points in P T -Symmetric Waveguides.
% - Physical review letters, 2008 - APS
%Klaiman et al. [
\emph{Phys. Rev. Lett.} \textbf{2008}, \emph{101}, 080402.


\bibitem{denisd}
{Znojil, M.}
Solvable model of quantum phase transitions
and the symbolic-manipulation-based study of its multiply degenerate exceptional
points and of their unfolding.
\emph{Ann. Phys.} \textbf{2013}, \emph{336}, 98--111.
%http://dx.doi.org/10.1016/j.aop.2013.05.016 (arXiv:1305.4822 [quant-ph])



\bibitem{denise}
Borisov, D.I.
Eigenvalues collision for PT-symmetric waveguide.
\emph{Acta Polytech.} \textbf{2014}, \emph{54}, 93.
%%(arXiv:1401.6316).

\bibitem{denisf}
Teimourpour, M.H.; Zhong, Q.; Khajavikhan, M.; El-Ganainy, R. Higher
Order EPs in Discrete Photonic Platforms. In \emph{Parity-Time
Symmetry and Its Applications}; Christodoulides, D., Yang, J.-K.,
Eds.;
%Hardback Springer Tracts in Modern Physics English
%Edited by  Demetrios Christodoulides ,
%Edited by  Jianke Yang
Springer: Singapore, 2018;
 pp. 261--276.

\bibitem{denisg}
Goldberg, A.Z.; Al-Qasimi, A.; Mumford, J.; O'Dell, D.H.J.
{Emergence of singularities from decoherence: Quantum catastrophes.}  \emph{Phys.
Rev. A} \textbf{2019}, \emph{100},  063628.


\bibitem{denish}
{Ramirez, R.; Reboiro, M.; Tielas, D.}
Exceptional Points from the Hamiltonian of a hybrid physical system:
Squeezing and anti-Squeezing.
\emph{Eur. Phys. J. D} \textbf{2020}, \emph{74}, 193.

\bibitem{PRSA}
Znojil, M.
Passage through exceptional point: Case study.
\emph{Proc. Roy. Soc. A: Math. Phys.
Eng. Sci.}
 {\bf 2020},
 {\em 476},
      20190831.


\bibitem{Uwe}
Guenther, U.; Stefani. F.
IR-truncated PT -symmetric $ix^3$ model and its asymptotic
spectral scaling graph. arXiv 1901.08526.


\bibitem{Uweb}
Semor\'{a}dov\'{a}, I.; Siegl, P.
Diverging eigenvalues in domain truncations of Schroedinger
operators with complex potentials.

2022 \emph{SIAM J. Math. Anal.} \textbf{2022}, in print
(arXiv:2107.10557).


\bibitem{catast}
Znojil, M.  {Quantum catastrophes: a case study.}
\emph{J. Phys. A Math. Theor.} \textbf{2012}, \emph{45},  444036.



\bibitem{Zeeman}
Zeeman, E.C.   {\em Cxatastrophe Theory-Selected Papers 1972--1977};
Addison-Wesley: Reading, UK, 1977.


\bibitem{Zeemanb}
Arnold, V.I.   \emph{Catastrophe Theory};  Springer: Berlin, Germany,  1992.

\bibitem{alliKG}
Mostafazadeh, A.
Hilbert space structures on the solution space of Klein-Gordon
type evolution equations.
\emph{Class. Quant. Grav.}
 {\bf 2003},
 {\em 20},
     155--171.


\bibitem{jaKG}
Znojil, M.
Relativistic supersymmetric quantum mechanics based on Klein-Gordon equation.
%J. Phys. A: Math. Gen. 37, 9557 (2004);
% - 9571.
%\bibitem{WDWja}
%
%Znojil, M.
%Relativistic supersymmetric quantum mechanics based on Klein--Gordon equation
%(hep-th/0408232).
%\emph{J. Phys. A Math. Gen.}
% {\bf 2004},
% {\em 37},
% 9557--9571.
%
%\emph{}
% {\bf },
% {\em },
%     -- .
%
%Znojil M. Solvable relativistic quantum dots with vibrational
%spectra. Czech J Phys 2005;55:1187–1192;
%
%M. Znojil, %Solvable relativistic quantum dots with vibrational
%%spectra. (quant-ph/0506017)
%Czech. J. Phys. 55, 1187 (2005).
%% - 1192
\emph{J. Phys. A: Math. Gen.}
 {\bf 2004},
 {\em 37},
     9557--9571.


\bibitem{aliWDW}
%
%\bibitem{aliKG}
Mostafazadeh, A.
Quantum mechanics of Klein-Gordon-type fields and quantum cosmology.
\emph{Ann. Phys. (N.Y.)}
 {\bf 2004},
 {\em 309},
     1--48.




\bibitem{WDW}
Znojil, M.
%2022
%Miloslav Znojil,
Wheeler-DeWitt equation
and the applicability of crypto-Hermitian interaction representation
in quantum cosmology.
\emph{Universe} \textbf{2022}, \emph{8},
 385.
%part of Special Issue
%Selected Topics in Gravity, Field Theory and Quantum Mechanics.
%DOI:10.3390/universe8070385
%(arXiv:2207.08508)


\bibitem{Denis}
Borisov, D.I.; R\r{u}\v{z}i\v{c}ka, F.; Znojil, M. Multiply
Degenerate Exceptional Points and Quantum Phase Transitions.
\emph{Int. J. Theor. Phys.} \textbf{2015}, \emph{54}, 42934305.

\bibitem{Denisb}
Znojil, M.; Borisov, D. I. Anomalous mechanisms of the loss of
observability
 in non-Hermitian quantum models.
\emph{Nucl. Phys. B} \textbf{2020}, \emph{957}, 115064.
%
%%\bibitem{anomalous}
%M. Znojil and D. I. Borisov,
%%Miloslav Znojil and Denis I. Borisov,
%%
% Nucl. Phys. B 957 (2020) 115064.
%%OPEN ACCESS  DOI: 10.1016/j.nuclphysb.2020.115064
%% (arXiv:2005.13069)


\bibitem{Denisc}
%\bibitem{ArnoldII}
Znojil, M.; Borisov, D.I.
Arnold's potentials and quantum catastrophes II.
\emph{Ann. Phys.} \textbf{2022},  \emph{442},  168896.
%DOI:10.1016/j.aop.2022.168896
%(arXiv:2101.02015v2)

\bibitem{Simon}
Eremenko, A.; Gabrielov, A.
{Analytic continuation of eigenvalues of a quartic oscillator.}
\emph{Comm. Math. Phys.} \textbf{2009}, \emph{287}, 431.

\bibitem{BenderWu}
{Bender,}
 C. M.; Wu, T. T. {Anharmonic oscillator.}
 \emph{Phys. Rev.} \textbf{1969}, \emph{184},  1231.


\bibitem{BB}
{Bender, C. M.; Boettcher, S.}
Real Spectra in Non-Hermitian Hamiltonians Having
PT Symmetry.
\emph{Phys. Rev. Lett.}
\textbf{1998}, \emph{80}, 5243.
%Bender C M and Boettcher S 1998 Phys. Rev. Lett. 80 5243 - 5246
%
%2009  Bender CM, Besseghir K, Jones HF, Yin X. Small-? behavior of
%the non-Hermitian PT-symmetric
%Hamiltonian H = p2 + x2(ix)? Journal of Physics a: Mathematical
%and Theoretical. 42.
%DOI: 10.1088/1751-8113/42/35/355301


\bibitem{Siegl}
%\bibitem{[3]}
Siegl, P.; Krej\v{c}i\v{r}\'{\i}k, D.
On the metric operator for the imaginary cubic oscillator.
%2012 Phys Rev D 86 121702(R)
%%(2012).
\emph{Phys. Rev. D} \textbf{2012}, \emph{86}, 121702(R).

\bibitem{Caliceti}
Caliceti, E.; Graffi, S.; Maioli, M.
\emph{Commun. Math. Phys.} \textbf{1980}, \emph{75},
51.

\bibitem{DB}
Bessis, D., private communication (1992).


\bibitem{BenderWub}
%Alvarez, G. {Bender-Wu branch points in the cubic oscillator.}
%\emph{J. Phys. A Math. Gen.} \textbf{1995}, \emph{27}, 4589.
{Alvarez, G.} Bender-Wu branch points in the cubic oscillator.
\emph{J. Phys. A Math. Gen.} \textbf{1995}, \emph{28}, 4589--4598.
%Published 21 August 1995 �
%Journal of Physics A: Mathematical and General, Volume 28, Number 16


\bibitem{SIGMA}
%M. Znojil, SIGMA 5, 001 (2009) (e-print overlay: arXiv:0901.0700).
{Znojil, M. Three-Hilbert-space} formulation of Quantum Mechanics.
\emph{Symm. Integ. Geom. Meth. Appl. SIGMA} {\bf 2009}, {\em 5}, 001.
%
%\bibitem{SIGMA}
%%M. Znojil, SIGMA 5, 001 (2009) (e-print overlay: arXiv:0901.0700).
%Znojil M 2009
%%. Three-Hilbert-space} formulation of Quantum Mechanics.
%%Symm. Integ. Geom. Meth. Appl.
%SIGMA 5 001


\bibitem{ozky}
Alase, A.; Karuvade, S.; Scandolo, C. M.
% 2022 J. Phys. A: Math. Theor.
%55 244003
The operational foundations
of PT-symmetric and quasi-Hermitian
quantum theory.
\emph{J. Phys. A: Math. Theor.} \textbf{2022}, \emph{55}, 244003.


\bibitem{Jonesb}
Jones, H. F. Interface between Hermitian and non-Hermitian Hamiltonians
in a model calculation.
%Physical Review D - Particles, Fields, Gravitation and Cosmology. 78.
%DOI: 10.1103/Physrevd.78.065032
%
%
\emph{Phys. Rev, D} \textbf{2008}, \emph{78}, 065032.
%
%arXiv:0805.1656  [pdf, ps, other]  hep-th quant-ph
%doi
%10.1103/PhysRevD.78.065032
%Interface between Hermitian and non-Hermitian Hamiltonians in a model calculation
%
%Authors: H. F. Jones
%
%Abstract: We consider the interaction between the Hermitian world,
%represented by a real delta-function potential -??(x), and the non-Hermitian
%world, represented by a PT-symmetric pair of delta functions with imaginary
%coefficients iß(?(x-L)-?(x+L)). In the context of standard quantum mechanics,
%the effect of the introduction of the imaginary delta functions on the bound-state
%energy of the real delt… ? More
%Submitted 15 September, 2008; v1 submitted 12 May, 2008;
%
%Comments: In this revised version we use a single metric operator for both
%bound and scattering states. The Discussion is amplified and extended
%
%Journal ref: Phys.Rev.D78:065032,2008


\bibitem{smeared}
Znojil, M. Scattering theory using smeared non-Hermitian
potentials.
%
\emph{Phys. Rev. D} \textbf{2009}, \emph{80},
045009.
%
%


\bibitem{sqwEP}
{Znojil, M.} Exceptional points and domains of unitarity for a class
of strongly non-Hermitian real-matrix Hamiltonians. \emph{J. Math.
Phys.} \textbf{2021}, \emph{62}, 052103.
%DOI:10.1063/5.0041185
%(arXiv:2104.11016)

\bibitem{maximal}
 Znojil, M. Maximal couplings in PT-symmetric chain-models
with the real spectrum of energies.
%
\emph{J. Phys. A Math. Theor.} \textbf{2007}, \emph{40},
4863 --
4875.
%
%





\bibitem{tridiagonal}
 Znojil, M. Tridiagonal PT-symmetric \emph{N} by \emph{N} Hamiltonians and a
fine-tuning of their observability domains in the strongly
non-Hermitian regime.
%
\emph{J. Phys. A Math. Theor.} \textbf{2007}, \emph{40},
13131--13148.
%
%


 \bibitem{BG}
Buslaev, V.; Grecchi, V. {Equivalence of unstable anharmonic
oscillators and double wells.} \emph{J. Phys. A Math. Gen.}
\textbf{1993}, \emph{26}, 5541--5549.

\bibitem{AKbook}
Albeverio, S.; Kuzhel, S.
PT-symmetric operators in quantum mechanics:
Krein spaces methods.
In
Bagarello, F.; Gazeau, J.-P.; Szafraniec, F.; Znojil, M. (Eds.)
\emph{Non-Selfadjoint Operators in Quantum Physics: Mathematical Aspects};
Wiley: {Hoboken, NJ, USA,}
2015.





\bibitem{Maple}
Char, B. W. et al, \emph{Maple V}; Springer-Verlag, New York, 1991.


\bibitem{without}
Znojil, M.
Unitary unfoldings of Bose-Hubbard exceptional
point with and without particle number conservation.
\emph{Proc.
Roy. Soc. A: Math., Phys. \& Eng. Sci. A} {\bf 2020},
 {\em 476},
 % 476 (2242)
%(2020)
20200292.


\bibitem{admissible}
{Znojil, M.} Admissible perturbations and false instabilities in
PT-symmetric quantum systems. %� PHYSICAL REVIEW A Volume: 97
%Issue: 3 Article Number: 032114 Published: MAR 16 2018
\emph{Phys. Rev. A} \textbf{2018}, \emph{97}, 032114.


\bibitem{admissibleb}
Znojil, M. Perturbation Theory Near Degenerate Exceptional Points.
 \emph{Symmetry} \textbf{2020}, \emph{12}, 1309.
%; section:
%
%DOI: 10.3390/sym12081309.
% (special issue
%Symmetries in Quantum Mechanics and Statistical Physics, G. Junker, Ed.)
%(arXiv:2008.00479)



\bibitem{corridors}
{Znojil, M.} Unitarity corridors to exceptional points.
\emph{Phys. Rev. A} \textbf{2019}, \emph{100}, 032124.
%DOI:10.1063/5.0041185
%(arXiv:2104.11016)


\bibitem{Christodoulides}
Christodoulides, D.; Yang, J.-K. (Eds.)
\emph{Parity-Time Symmetry and Its Applications};
Springer: Singapore, 2018.

\bibitem{SKbook}
Krej\v{c}i\v{r}\'{\i}k, D.; Siegl, P.
Elements of spectral theory without the spectral theorem.
In
Bagarello, F.; Gazeau, J.-P.; Szafraniec, F.; Znojil, M. (Eds.)
\emph{Non-Selfadjoint Operators in Quantum Physics: Mathematical Aspects};
Wiley: {Hoboken, NJ, USA,}
2015.

\bibitem{PSPC}
Siegl, P., private communication (2016).


\end{thebibliography}
\end{document}